\documentclass{article}

\usepackage{microtype}
\usepackage{subfigure}
\usepackage{booktabs} 
\usepackage{hyperref}
\usepackage{times}

\usepackage{amsmath,amsfonts,bm}

\def\eqref#1{equation~\ref{#1}}

\def\1{\bm{1}}

\DeclareMathAlphabet{\mathsfit}{\encodingdefault}{\sfdefault}{m}{sl}
\SetMathAlphabet{\mathsfit}{bold}{\encodingdefault}{\sfdefault}{bx}{n}

\DeclareMathOperator*{\argmax}{arg\,max}

\usepackage{hyperref}
\usepackage{url}
\usepackage[utf8]{inputenc} 
\usepackage[T1]{fontenc}    
\usepackage{amsfonts}       
\usepackage{amssymb}
\usepackage{pifont}
\newcommand{\xmark}{\ding{55}}%
\usepackage{nicefrac}       
\usepackage{microtype}      
\usepackage{xcolor}         
\usepackage[algo2e]{algorithm2e} 
\usepackage{enumitem}
\usepackage{makecell}
\usepackage{multirow}
\usepackage[bottom]{footmisc}
\usepackage{graphicx}
\usepackage{stfloats}
\usepackage{float}
\newcommand{\myname}{MIP\xspace}

\newcommand{\nop}[1]{}

\usepackage[accepted]{icml2023}

\usepackage{amsmath}
\usepackage{amssymb}
\usepackage{mathtools}
\usepackage{amsthm}
\usepackage{bbm}
\usepackage{bm}

\usepackage[capitalize,noabbrev]{cleveref}

\theoremstyle{plain}

\theoremstyle{definition}

\theoremstyle{remark}

\usepackage[textsize=tiny]{todonotes}

\icmltitlerunning{Everyone's Preference Changes Differently: A Weighted Multi-Interest Model for Retrieval}

\begin{document}

\twocolumn[
\icmltitle{Everyone's Preference Changes Differently: \\ A Weighted Multi-Interest Model for Retrieval}



\icmlsetsymbol{equal}{*}

\begin{icmlauthorlist}
\icmlauthor{Hui Shi}{yyy}
\icmlauthor{Yupeng Gu}{comp}
\icmlauthor{Yitong Zhou}{comp}
\icmlauthor{Bo Zhao}{comp}
\icmlauthor{Sicun Gao}{yyy}
\icmlauthor{Jishen Zhao}{yyy}
\end{icmlauthorlist}

\icmlaffiliation{yyy}{Department of Computer Science and Engineering, University of California San Diego, La Jolla, United States, \textit{\{hshi, jzhao, sicung\}@ucsd.edu}}
\icmlaffiliation{comp}{Pinterest, San Francisco, United States, \{yupeng, yzhou, bozhao\}@pinterest.com}

\icmlcorrespondingauthor{Hui Shi}{hshi@ucsd.edu}


\icmlkeywords{Recommendation System}

\vskip 0.3in
]



\printAffiliationsAndNotice{}  

\begin{abstract}


User embeddings (vectorized representations of a user) are essential in recommendation systems. Numerous approaches have been proposed to construct a representation for the user in order to find similar items for retrieval tasks, and they have been proven effective in industrial recommendation systems. Recently people have discovered the power of using multiple embeddings to represent a user, with the hope that each embedding represents the user's interest in a certain topic. With multi-interest representation, it's important to model the user's preference over the different topics and how the preference changes with time. However, existing approaches either fail to estimate the user's affinity to each interest or unreasonably assume every interest of every user fades at an equal rate with time, thus hurting the performance of candidate retrieval. In this paper, we propose the Multi-Interest Preference (\myname) model, an approach that not only produces multi-interest for users by using the user's sequential engagement more effectively but also automatically learns a set of weights to represent the preference over each embedding so that the candidates can be retrieved from each interest proportionally. Extensive experiments have been done on various industrial-scale datasets to demonstrate the effectiveness of our approach. 
\footnote{The code is available at: \url{https://github.com/shihui2010/MIP}}
\end{abstract}

\section{Introduction}
\nop{
outline:
- recommendation/retrieval problem
- user representation and retrieval (candidate generation) in recsys
- embedding-based approach
- embedding-based approach + sequence modeling
///
- multi-interest
- multi-interest + weight

}

\begin{figure*}[pt]
    \centering
    \includegraphics[width=0.8\textwidth]{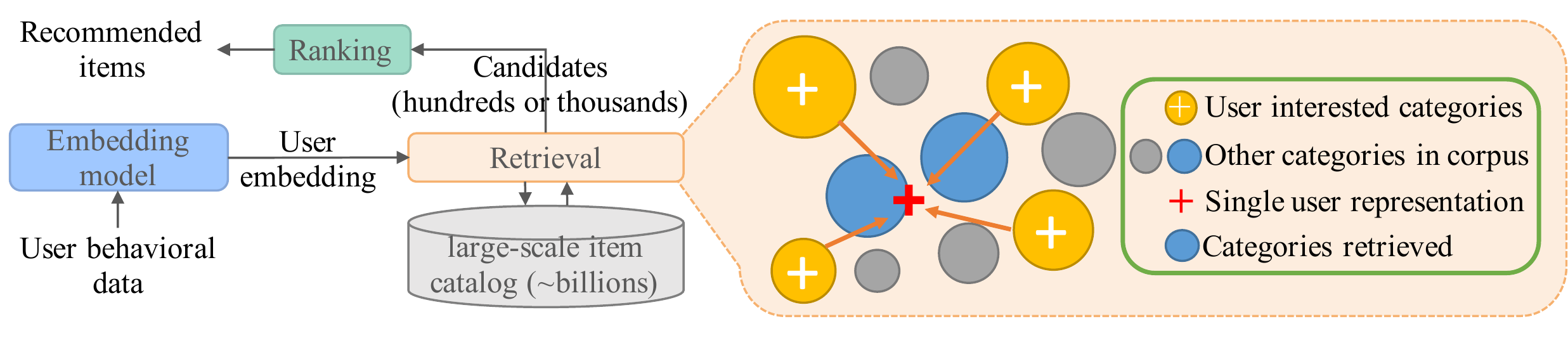}
    \vspace{-0.3in}
    \caption{Mis-representation with single user embedding in the retrieve-then-rank framework.
    }
    \label{fig:multi_interest_motivation}
\end{figure*}

Today, the recommendation system is widely used in online platforms to help users discover relevant items and deliver a positive user experience. In industrial recommendation systems, there are usually billions of entries in the item catalog, which makes it impossible to calculate the similarity between a user and every item. The common approach is, illustrated in Figure~\ref{fig:multi_interest_motivation}, retrieving only hundreds or thousands of candidate items based on their similarity to the user embedding on an approximate level (\emph{e.g.} inverted indexes, locality-sensitive hashing) without consuming too much computational power, and then sending the retrieved candidates to the more nuanced ranking models.
Thus, finding effective user embedding is fundamental to the recommendation quality.

The user representations learned from the neural networks are proven to work well on large-scale online platforms, such as Google \citep{cheng2016wide}, YouTube \citep{covington2016deep}, and Alibaba \citep{wang2018billion}. 
Mostly, the user embeddings are learned by aggregating the item embeddings from the user engagement history, via sequential models \cite{gru4rec,quadrana2017personalizing,kang2018self,you2019hierarchical}. These works usually rely on the sequential model, \emph{e.g.} a Recurrent Neural Network (RNN) model or an attention mechanism, to produce a single embedding that summarizes the user's one or more interests from recent and former actions. 

Recently researchers \citep{epasto2019single, weston2013nonlinear, pinnersage, li2005hybrid} have discovered the importance of having multiple embeddings for an individual, especially in the retrieval phase, with the hope that they can capture a user's multiple interests. The intuition is quite clear: if multiple interests of a user are collapsed into a single embedding, though this embedding could be similar to and can be decoded to all the true interests of the user, directly using the single collapsed embedding to retrieve the closest items might result in items that the user is not quite interested in, as illustrated in Figure~\ref{fig:multi_interest_motivation}.

Though, conventional sequential models like RNN or the Transformer network do not naturally produce multiple sequence-level embeddings as desired in the multi-interest user representation. Existing solutions fall into two directions: 1) \textbf{split-by-cluster} approaches first cluster the items in the user engagement history by category labels \citep{li2019multi} or item embedding vectors \citep{pinnersage} and then compute a representation embedding per cluster; 2) \textbf{split-by-attention} models adopt transformer-like architecture with two modifications. The query vectors in the attention are learnable vectors instead of the projections from the input and the results of each attention head are directly taken as multiple embeddings \citep{pintext2, comirec}. The limitations of the two approaches are obvious: the split-by-cluster method works best with dense item feature \citep{xue2005scalable}; and split-by-attention models bias towards the popular categories owing to its shared query vector among all the users and are inflexible to adjust the number of interests, which is fixed in the training phase as the number of attention heads. 

Moreover, the existing multi-interest works ignore one important aspect: the weights of each embedding. 
In the retrieval stage, given the limited number of items to return, retrieving items from each embedding uniformly will cause a recall problem when the user clearly indicates a high affinity towards one or two categories. Some existing approaches, \emph{e.g.} PinnerSage \citep{pinnersage}, use exponentially decayed weights to assign a higher score to interests that have more frequent and recent engagements. However, the methods still assume that in the same period, regardless of whether the interest is enduring or ephemeral, the level of interest decays equally for any user. Furthermore, these works also assume the number of embeddings to be fixed across all users. Not only is this hyperparameter costly to find, but also the assumption that all users have the same number of interests is questionable. Some dormant users can be well represented using one or two vectors, while others might have a far more diverse set of niche interests that requires tens of embeddings to represent. 

In this paper, we propose Multi-Interest Preference (\myname) model that learns user embeddings on multiple interest dimensions with personalized and context-aware interest weights. The \myname model consists of a clustering-enhanced multi-head attention module to compute multiple interests and a feed-forward network to predict the weights for each embedding from the interest embedding as well as the temporal patterns of the interest. The clustering-enhanced attention overcomes the aforementioned shortcomings from two aspects: the query, key, and value vectors are projected from the user's engaged items, thus the output of the attention is personalized and minimized the bias toward globally popular categories; moreover, the clustering module can be applied before or after the multi-head attention, releasing the assumption that item features are pre-computed or the item-category labels are available. 
The main contribution of this paper and the experimental evidence can be summarized as follows:
\begin{itemize}[noitemsep, topsep=0pt]
    \item We propose a multi-interest user representation model that minimizes the bias towards popular categories and is applicable no matter if the item embeddings are pre-computed. \myname is successful in various industry-scale datasets (Section \ref{sec:cf_exp}, \ref{sec:pin_exp}); Appendix~\ref{ssec:syn_exp} reveals the bias resulting from the global query vector and the error resulting from a fixed number of clusters in the split-by-attention approaches, in comparison to \myname. 
    \item In addition to the multi-facet vector representations of a user, \myname assigns weights to each embedding, which are automatically customized for each user interest. This approach improves the recall of candidate generation by retrieving more candidates from the most representative embedding. (Section~\ref{sec:ablation_study}). 
    \item Although if the clustering algorithms require, \myname still asks for a number of clusters during the training phase, the number of clusters in \myname in the inference phase can be trivially increased or decreased without re-training of the model. And the experimental results (Appendix~\ref{ssec:abl_cluster}) show that re-configuring the number of clusters has an insignificant impact on the retrieval performance, thus allowing the system to trade off the storage and computation cost for better performance. Thus, \myname does not require prior knowledge of the number of interests of users during the model training phase.
\end{itemize}

\label{sec:intro}

\section{Related Work}
\begin{table*}
\centering
\begin{tabular}{c|cccc}
\hline\hline 
   Name  & \makecell{User \\Embedding} & \makecell{Sequential \\Model} & \makecell{Additional\\ Input} & \makecell{Preference \\Weight}\\   
   \hline 
   GRU4Rec \cite{gru4rec} & single & RNN & interaction session & N/A \\
   TiSASRec \cite{tisas} & single & time-aware self-attention & timestamps & N/A\\
   BERT4Rec \cite{sun2019bert4rec} & single & self-attention & -- & N/A\\
   MIND\cite{li2019multi} & multiple & label-aware self-attention & category labels & \xmark\\
   ComiRec \cite{comirec} & multiple & global-query attention & -- & \xmark\\
   PinText2 \cite{pintext2} & multiple & shared global-query attention & -- & \xmark\\
   PinnerSage \cite{pinnersage} & multiple & N/A & -- & heuristic \\ 
   \hline
   \myname & multiple & time-aware self-attention & timestamps & learned \\
   \hline 
\end{tabular}
\caption{Comparison of \myname to existing recommendation models.}
   \label{tab:related_work_compare}
\end{table*}

This work relates to two important aspects of existing recommendation systems: sequential models and the multi-interest framework. 

\textbf{Sequential models.} A basic consensus in the recommendation system is that user embeddings should be inferred from the user's historical behavior, and thus the sequential models have been at the heart of recommendation models. A typical and classical sequential model is the Markov Chain \citep{rendle2010factorizing, he2016fusing}. While Markov Chain captures short-term patterns of engagement sequence well, it fails to make the recommendation that requires memorizing long sequences. With stronger representation power on long sequences, Recurrent Neural Networks (RNNs) have been adopted for learning user embedding from arbitrarily long sequences, \emph{e.g.} GRU4Rec \citep{gru4rec} and others \citep{xu2019recurrent, devooght2017long}. 
Besides the standard RNN models, specialized recurrent units are proposed to meet the special need of incorporating certain information, \emph{e.g.} user demographic information \citep{donkers2017sequential}, global context \citep{xia2017attention}, interest drifts with time \citep{chen2019dynamic}, and interaction session \citep{gru4rec}. Recently, the success of the Transformer network \citep{vaswani2017attention} has brought revolution to sequential modeling tasks\cite{shi2022learning, perez2019turing} and has been soon adapted to the recommendation models, \emph{e.g.} ComiRec \citep{comirec}, BERT4Rec \citep{sun2019bert4rec}, TiSASRec \citep{tisas}, SASRec \citep{kang2018self}, MIND \citep{li2019multi}, PinText2 \citep{pintext2}, and also our \myname.


\textbf{Multi-interest user representation.}
Representing users by multiple embeddings greatly improves the recommendation quality, but not every existing recommendation model can easily extend to a multi-interest framework. Classical collaborative filtering and matrix factorization methods do not naturally produce multiple user embeddings, and so do RNNs and attention-based models. To discover multiple interests from user engagement history, heuristic methods \citep{jiang2020aspect, yue2012multi} and unsupervised learning methods like clustering \citep{pinnersage, wandabwa2020multi} and community mining \citep{wang2007multi, yu2008using} have been adopted. Besides, researchers have made efforts to modify the existing neural networks to produce multiple results, for instance, the capsule network \citep{li2019multi, sabour2017dynamic, comirec} and multi-head attention models \citep{tisas, comirec, zhou2018atrank}. However, they require an estimation of the number of interests of users as a hyperparameter and do not learn the weight of interests. Therefore, unlike \myname, they produce an equal number of clusters for every user and treat each interest with uniform importance. 

\textbf{Relationship to previous works.} The motivation of \myname is to acquire weighted multiple user embeddings with standard self-attention but without explicit item-category labels. ComiRec and PinText2 use global-query attention to produce multiple embeddings, which introduces a bias toward frequent items or popular categories and the phenomenon is shown in Appendix~\ref{ssec:syn_exp}. Furthermore, they also predefine a number of interests that is uniform for all the users. TiSASRec and BERT4Rec adopt self-attention but can not learn multiple embeddings. MIND relies on the category labels to produce multiple embeddings from self-attention and capsule networks. However the category labels are sometimes unavailable or vague in other applications, \emph{e.g.} YouTube and Pinterest. PinnerSage produces multiple embeddings without category labels, but requires pre-computed item embeddings. The comparison are summarized in Table~\ref{tab:related_work_compare}.

\label{sec:relate}

\section{Methodology}

\nop{

User Representation:
-- Item Encoding
-- -- item embedding
-- -- Temporal/Positional Encoding with multiple options
-- Self-Attention
-- -- cluster assignment with exogenous item feature vector
-- -- cluster assignment without item feature vector
-- multi-interest representation

Cluster Weight learning:
-- cluster weights....

User-Item Engagement Prediction

Training:
}

In this section, we formulate the recommendation problem and the neural architecture to model the multiple user interests with preference weights in detail. 

\subsection{Problem Statement}
\label{ssec:method_problem}


Let $\mathcal{I}$ denote the collection of items and $\mathcal{U}$ denote the set of users. The interaction sequence of a user $u \in \mathcal{U}$ is represented as $\mathcal{S}^u$ with a list of item IDs $(v_{t_1}^u, v_{t_2}^u,..., v_{t_{l_u}}^u)$ and timestamps $(t_1^u, t_2^u, ..., t_{|l_u|}^u)$. Each item $v_i^u \in \mathcal{I}$ is associated with an item embedding $\bm{p}_i^u \in \mathbb{R}^d$ and $l_u$ is the length of the interaction sequence.

The objective is to learn a set of user embeddings $\bm{z}_\lambda^u \in \mathbb{R}^d$ and their weights $w_\lambda^u$ ($\lambda=1, ..., \Lambda$) for each user $u$. Since the user representation is learned only from the history of that user, hereinafter, we omit the superscript for $u$ in both the input and output sides for simplicity. Notations are summarized in Table~\ref{tab:notations}.

Item embedding $\bm{p}_j$ can be either represented by item metadata features or treated as model parameters and learned from the data. When items have dense metadata features (\emph{e.g.} text embeddings), we will use them as the item embedding and focus on learning the user representation. When items are only represented by an ID and do not have other metadata, we will learn the item embedding table. 
Our model is able to handle both cases and eventually learn users' multi-interest embeddings and their weights.

\begin{table}
    \centering
    \resizebox{\columnwidth}{!}{
    \begin{tabular}{l|l}
    \hline\hline
       Notations  &  Description\\
       \hline
        $\mathcal{I}, \mathcal{U} $ & Item set and user set \\ 
        $\mathcal{S}$ & Abbreviation of user engagement history $\mathcal{S}^u$ \\
        $l$ & Abbreviation of $l_u$, length of $\mathcal{S}$ \\
        $d$ & The item embedding dimension \\
        $d_{model}$ & Projected key/query vector dimension \\
        $h$ & Attention head superscript \\
        $\bm{p}$ & An item embedding, $\bm{p} \in \mathbb{R}^d$ \\
        $\bm{p}_j, t_j$ & Short form of the $\bm{p}_{t_j}^u$ and $t_j^u$ \\
        $W^h_q, \bm{b}^h_q$ & Query projection weights and bias \\
        $W^h_k, \bm{b}^h_k$ & Key projection weights and bias \\
        $\bm{M}$ & Attention mask matrix \\
        $\Lambda$ & Maximum number of interests per user\\
        $\mathcal{C}$ & \makecell[l]{Cluster assignment, $\mathcal{C}\in \mathbb{R}^l$. \\ $\mathcal{C}_{[i]}=\lambda$ if i-th item belongs to $\lambda$-th cluster} \\
        $\bm{L}_\lambda$ & The set of indices of items in the $\lambda$-th cluster. \\
        $\bm{z}_\lambda, Z$ & User embedding vector(s)\\ 
        $\mathbbm{1}_{[cond]}$ & 1 if $cond$ is true, 0 otherwise \\
        $\bm{\tau}_{[m]}$ & m-th digit in the vector $\bm{\tau}$ \\
        $[;]$ & Vector concatenation operator  \\
        \hline
    \end{tabular}
    }
    \caption{Notations}
    \label{tab:notations}
    \vspace{-0.3in}
\end{table}




\begin{figure*}[t]
    \centering
    \includegraphics[width=5.5in]{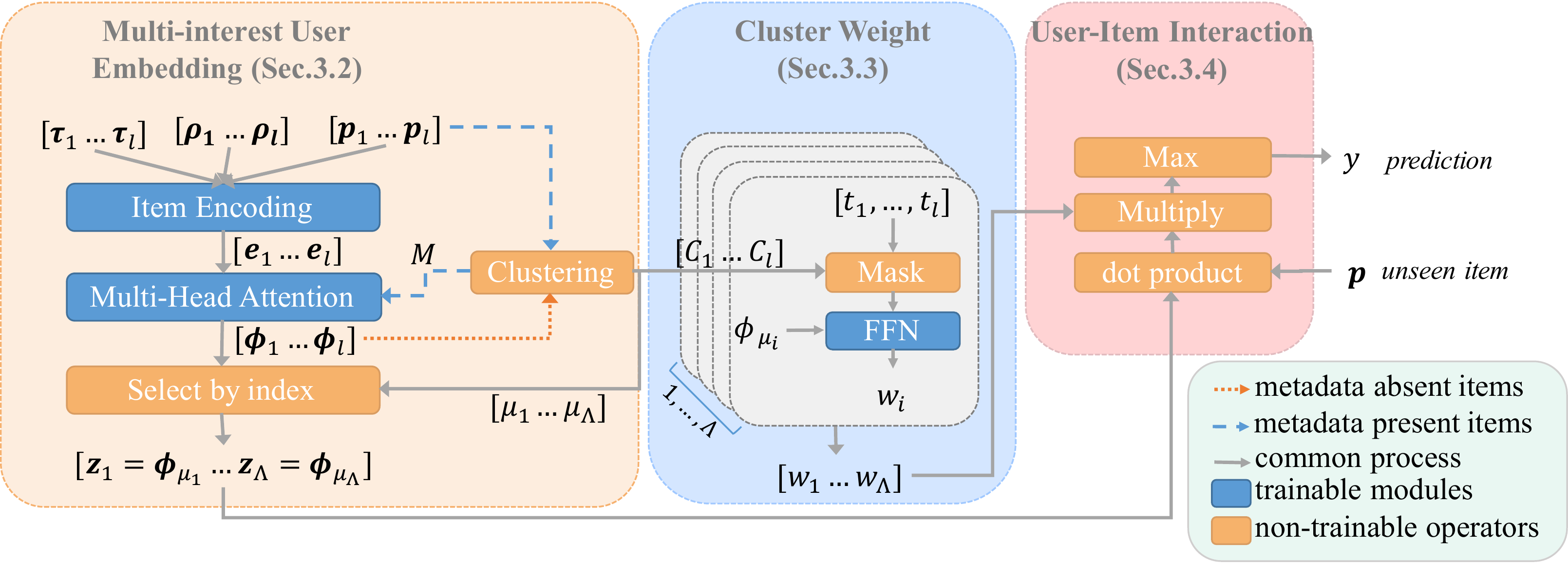}
    \caption{An overview of \myname architecture. The input is the user engagement history containing item embeddings $[\bm{p}_1 ... \bm{p}_l]$), temporal encoding $[\bm{\tau}_1...\bm{\tau}_l]$, and positional encoding ($[\bm{\rho}_1...\bm{\rho}_l]$). The multi-interest user embedding module produces $\Lambda$ embeddings, where $\Lambda$ is decided by the clustering method or as a hyperparameter. With clustering and multi-interest representation, the cluster weight module will then estimate the cluster weights for each cluster. Finally, the multi-interest embeddings with corresponding weights are combined to predict the user's interest in an unseen item $\bm{p}$. The processes with item metadata absent/present are shown with different arrows.}
    \label{fig:overview}
\end{figure*}

\subsection{Multi-Interest User Modeling}
\label{ssec:method_multi_interest}

\textbf{Item Representation.}
\nop{
We distinguish two conditions of an item. In most practical systems, every item has a pre-computed feature vector (denote as $\bm{p}_j$) defined in the space where the distance can capture the similarity between the items. In other cases where the recommendation system knows nothing but unique ids (denoted by an one-hot vector $\bm{v}_j$) about the item, the system needs to learn the dense item feature through an embedding layer:
\begin{equation}
    \small
    \bm{p}_j = W_{emb} \bm{v}_j
    \label{eq:item_emb}
\end{equation}
}
For the j-th item in the user engagement history, we represent the item by $\bm{p}_j$ and it's either learned (when item metadata is absent) or copied from the item metadata feature (when item metadata is present).
In addition to $\bm{p}_j$, the sequential order and relative timestamps of the interactions are represented by positional encoding and temporal encoding respectively as $\bm{\rho}$ and $\bm{\tau}$. Following \citet{vaswani2017attention}, the odd digits ($2m+1$) and even digits ($2m$) on the encoding vectors are given by:
\begin{equation}
\small
\begin{aligned}
  & \bm{\tau}_{[2m]}(t_j) = sin \big( t_j / (\tau_{max})^{2m / m_t} \big) \\
  &  \bm{\tau}_{[2m + 1]}(t_j) = cos \big( t_j / (\tau_{max})^{2m / m_t} \big) \\
  & \bm{\rho}_{[2m]}(j) = sin \big( j / (\rho_{max})^{2m / m_p} \big) \\
  &  \bm{\rho}_{[2m + 1]}(j) = cos \big( j / (\rho_{max})^{2m / m_p} \big)
\end{aligned}
\label{eq:encoding}
\end{equation}

The hyper-parameters are set as $\tau_{max}=\rho_{max}=1 \times 10^4$, and $m_t=m_p=1$. The timestamps unit is \textit{day}. In the Section~\ref{ssec:variants_positional_temporal_encoding}, other encodings forms are compared, and show that the design used in Equation~\ref{eq:encoding} has a slight advantage. 

Finally, item ID embedding is concatenated with the positional and temporal encodings to produce the final representation of an interaction:
\begin{equation}
    \mathbf{e}_j := [\bm{p}_j; \bm{\tau}(t_j); \bm{\rho}(j)]
    \label{eq:concat_encoding}
\end{equation}
\textbf{User Representation.}
Our user representation is built upon the multi-head self-attention module \cite{vaswani2017attention}. Using the interaction embedding as above, for each attention head $h$, we define the attention weight between interaction $i$ and $j$ as
\begin{equation}
\small
a^h_{i, j} = \frac{\exp (s^h_{i, j})}{\sum_{k=1}^{l} \exp (s^h_{i, k})}
\label{eq:att}
\end{equation}
where $s^h_{i, j}$ is the dot product between the projected query of $j$ and the projected key of $i$:
%
\begin{equation}
    \small
    \begin{aligned}
    & s^h_{i, j} = \Big( (W^h_q \bm{e}_j + \bm{b}^h_q)^{\top} \cdot (W^h_k \bm{e}_i + \bm{b}^h_k)
     \Big) / \sqrt{d_{model}}
    \end{aligned}
\end{equation}
%
%
In order to build the user's multi-interest embedding, we need to cluster the items in the user sequence. When we use item metadata features as $\bm{p}$, we  pre-compute the item clusters $L_1, \cdots, L_\Lambda$ where each $L_\lambda$ is the set of item IDs that belong to $\lambda$-th cluster. Let $\mathcal{C}$ denote the mapping from item ID to cluster assignment, \emph{i.e.} $\mathcal{C}_{[j]} = \lambda$ if $j \in L_\lambda$. Given the cluster information, we have the advantage of summarizing similar items into a single representation. Specifically, when the context vector only attends to items within the same cluster as the current item, we force that vector to contain only the information from that cluster. Naturally, a mask is introduced to enforce such constraint: let $M \in \{0,1\}^{l \times l}$ be the mask matrix where $M_{i,j} = \mathbbm{1}_{[\mathcal{C}_{[i]} = \mathcal{C}_{[j]}]}$ (and 0 otherwise). Each attention head $h$ produces the context vector at position $j$ by aggregating the sequence as:
%
\begin{equation}
    \small
    \bm{\phi}^h_j = \sum_{i=1}^{l}  a_{i,j}^h M_{i,j} \bm{p}_i
\label{eq:zh}
\end{equation}
To process the context vector from all attention heads, a dropout layer and a feed-forward network (FFN) are applied, and the output vector is computed as
\begin{equation}
\label{eq:compute_z}
    \small
    \bm{\phi}_j = FFN(Dropout([\bm{\phi}^1_j; ...;\bm{\phi}^H_j])), ~~~ j=1,\cdots,l
\end{equation}
The $FFN()$ consists of two fully-connected linear layers with a hyperbolic tangent activation function after the first layer, \emph{i.e.} $FFN(x)=W(tanh(W'x+b')) + b$.

When item metadata is absent and $\bm{p}$ need to be learned, the mask $M$ will be an all-ones matrix, and output vectors are still computed according to Equation \ref{eq:zh} and \ref{eq:compute_z}. Then the item clusters are computed from $\bm{\phi}$ instead of from $\bm{p}$. We still use $\mathcal{C}$ to represent such cluster assignment, and it will be used to define the user's multi-interest embedding as below.


So far, the multi-head attention module has produced $l$ vectors $\bm{\phi}_1, ..., \bm{\phi}_l$, and each $\bm{\phi}_j$ uses $\bm{p}_j$ as the (unprojected) query. We will build the multi-interest user embedding by selecting the $\Lambda$ context vectors that represent each cluster.
Denote the position of the last item in each cluster $\lambda$ as $\mu_\lambda$ (\emph{i.e.} $\mu_\lambda = \argmax_{j} (\mathcal{C}_{[j]}=\lambda) $), we will take context vector at that position 
to represent the $\lambda$-th user embedding. 
In sum, the multi-interest user embedding is
%
\begin{equation}
    \small
    Z = [\mathbf{z}_1^\top; ...; \mathbf{z}_\Lambda^\top]
    = [\bm{\phi}_{\mu_1}^\top; ...; \bm{\phi}_{\mu_\Lambda}^\top]
    \in \mathbb{R}^{\Lambda \times d}
\label{eq:z}
\end{equation}
Each $\bm{z}_\lambda$ attends to only items that belong to the same cluster as the item on position $\mu_\lambda$.


\subsection{Cluster Weight Modeling}
\label{ssec:method_weight}
Besides the multi-interest user embedding, it's also likely that a user favors each interest unequally. As mentioned earlier, ranking these interests correctly can greatly benefit the candidate generation task given its limited budget. PinnerSage \cite{pinnersage} uses an exponential-decay heuristic function to represent the weight for a cluster, following the assumption that more recent interactions should contribute more to the cluster weight. While we believe that the intuition is generally true, it would be best for the model to automatically learn the role of timestamps from the data. We also incorporate the user embeddings on that cluster's dimension ($\bm{z}_{\lambda}$) into cluster weight modeling, since certain categories of interest can have different impacts on the cluster weights as well. Therefore we design the cluster weights to be a function of the recency of the interaction and user embeddings.
%
%
%
We model the weight of cluster $\lambda$ as
\begin{equation}
    \small
    w_\lambda = FFN([\bm{z}_\lambda; ~ \mathbbm{1}_{[C_{[1]} = \lambda]} \cdot \bm{\tau}_{1};... ; \mathbbm{1}_{[C_{[l]} = L_\lambda]} \cdot \bm{\tau}_{l}])
    \label{eq:cluster_weight}
\end{equation}

%
The first part inside the $FFN$ is the user's embedding on cluster $\lambda$.
The second part inside the $FFN$ serves as a mask that retains only the timestamps of relevant items that belong to cluster $\lambda$.
We will discuss how to learn these weights in the next subsection.
%


\subsection{User-Item Interaction Modeling}
\label{ssec:latent_factor}

On the item level, intuitively, a user will engage with an item as long as the item matches \textit{at least one} of his/her interests (not \textit{all}). For example, a user who is interested in both running and home decor purchased a lawn mower, and this behavior will be explained by the user's embedding of the ``home decor'' cluster (\emph{i.e.} $\bm{z}_{home}$) and has nothing to do with the embedding of the ``running'' cluster (\emph{i.e.} $\bm{z}_{running}$). In other words, the similarity of $\bm{z}_{home}$ and $\bm{p}_{lawn\ mower}$ should be high, and the similarity of $\bm{z}_{running}$ and $\bm{p}_{lawn\ mower}$ should not even matter. Therefore, when measuring the user-item affinity, we should consider the one user embedding that is the most similar (\emph{e.g.} highest cosine similarity) to the item.

On the cluster level, we need another factor to explain a user's behavior towards different clusters. This can no longer be represented simply by the semantic similarity in the embedding space any more. When the user purchases 20 items in the ``home decor'' cluster and 5 items in the ``running'' cluster, it does not indicate that the similarities between $\bm{z}_{home}$ and these 20 items are higher than the similarities between $\bm{z}_{running}$ and these 5 items (in fact, all of the similarities should be as high as possible). Therefore, we will multiply the user-item affinity by the cluster weight here to represent a user's intensity towards different clusters. 

Considering the arguments above, we propose the likelihood that a user (represented by $Z = [\mathbf{z}_1^\top; ...; \mathbf{z}_\Lambda^\top]$) interacts with an item (represented by $\bm{p}$) as follow:
%
%
\begin{equation}
    \small
    y = \max \{w_\lambda \cdot (\bm{z}_\lambda \cdot \bm{p}) \}_{\lambda=1}^{\Lambda}
\label{eq:aggregator}
\end{equation}
%

Given the set of items with positive label ($\{ \mathcal{I}_+^u \}_{u \in \mathcal{U}}$) and negative label ($\{ \mathcal{I}_-^u \}_{u \in \mathcal{U}}$), the negative log-likelihood (NLL) loss of our model can be written as:
\begin{equation}
    \small
    \mathcal{L} = -
    \frac{\sum_{u \in \mathcal{U}} \Big(
        \sum_{\bm{p}_i \in \mathcal{I}^u_+} \log(y^u_i) +
        \sum_{\bm{p}_i \in \mathcal{I}^u_-} \log(1 - y^u_i)
        \Big) }
    {\sum_{u \in \mathcal{U}} \Big(
        |\mathcal{I}^u_+| + |\mathcal{I}^u_-| 
    \Big)}   
    \label{eq:nll}
\end{equation}

\label{sec:method}

\section{Experiments}
%

We conduct an exhaustive analysis to demonstrate the effectiveness of \myname on the data from Pinterest, one of the largest online content discovery platforms, and a few public datasets. We will divide our discussions to two categories: (1) learning item ID embeddings (Section~\ref{sec:cf_exp}) and (2) using item metadata features as is (Section~\ref{sec:pin_exp}). Finally, an ablation study is done on different components of the model (Section \ref{sec:ablation_study}).

\subsection{Learning Item ID Embeddings}
\label{sec:cf_exp}

\begin{table}
    \centering
    \resizebox{0.9\columnwidth}{!}{
    \begin{tabular}{l|ccc}
    \hline\hline
         &  Amazon & MovieLens & Taobao \\
    \hline
    \# Items & 425,582 & 15,243 & 823,971 \\
    \# Interactions & 51M & 20M & 100M \\
    \# Training Seq & 57,165 & 127,212 & 343,171 \\
    \# Test Seq & 5,000 & 5,000 & 10,000 \\
    \# Validation Seq & 5,000 & 5,000 & 10,000 \\
    \hline
    \end{tabular}
    }
    \caption{Dataset statistics.}
    \label{tab:datasets}
\end{table}

We first evaluate \myname on learning from collaborative filtering datasets, where the item features are absent and will be learned from the user-item interactions.

\textbf{Dataset.}
Three public datasets are used: Amazon-book \footnote{https://jmcauley.ucsd.edu/data/amazon/} (hereinafter, Amazon),
Taobao\footnote{https://tianchi.aliyun.com/dataset/dataDetail?dataId=649}, and 
MovieLens\footnote{https://www.kaggle.com/grouplens/movielens-20m-dataset}. We adopt a 10-core setting as previous works \citep{tisas, wang2019neural} and filter out items that appear less than 10 times in the dataset. 
We then split each user's engagement history to non-overlapping sequences of length 100, and use the first 50 items to learn the user embeddings and the last 50 items as labels (as used in \citet{comirec}). Any sequence shorter than this threshold are discarded. For each sequence, another 50 negative samples are uniformly sampled at random from the items that the user does not interact with. Our goal is to rank the positive items (that users have actually interacted with) higher than the negative items (random). The dataset statistics are listed in Table~\ref{tab:datasets}.

\textbf{Baseline and model configuration.} 
We compare several open-sourced baseline models with \myname. For fair comparison, we set up the configurations as follow: (1) item and user embedding vectors have the same size ($d=32$); (2) the number of attention heads is the same ($H=8$) if the model includes a multi-head attention module; (3) the baseline models should have similar or more parameters than \myname. We let the hidden size in GRU4Rec \cite{gru4rec} be 128, the key and query projected dimension ($d_{model}$ in Equation~\ref{eq:att}) is labeled in place with the results, and if the model contains a position-wise FFN (Equation~\ref{eq:compute_z}), it will be a two-layered fully-connected structure with a hidden size of 32 each. The BERT4Rec model \cite{sun2019bert4rec} is originally proposed to predict the item directly as a classification task, so we take its last BERT output as the user embedding to compute the similarity between user and item, and train with the NLL loss. We disabled the session-parallel mini-batch in these models since the session information is absent. We also replace the text encoder in the PinText2 \cite{pintext2} with an item embedding layer since the inputs in our experiments are items instead of texts. 

\textbf{Training setup.}
All the models are trained for 100 epochs on a NVIDIA Tesla T4 GPU with an early stop strategy that stops the training when validation AUC does not improve for 20 epochs. The clustering method used in \myname is the Ward clustering algorithm \citet{ward1963hierarchical}.\footnote{We adopted the scikit-learn implementation. \url{https://scikit-learn.org/stable/modules/generated/sklearn.cluster.AgglomerativeClustering.html}, with n\_cluster=5, and other default arguments. } We compare other clustering methods in Appendix~\ref{ssec:abl_cluster}.

\begin{table*}[]
    \centering
    \resizebox{\textwidth}{!}{
    \begin{tabular}{c|c|cc|ccc|ccc|ccc|c}
       \hline\hline
       \multirow{2}{*}{Model} & \multirow{2}{*}{Params.} & \multicolumn{2}{c|}{Config} & \multicolumn{3}{c|}{Amazon} & \multicolumn{3}{c|}{Taobao} & \multicolumn{3}{c|}{MovieLens}   & \multirow{2}{*}{\makecell{Latency\\(\textit{ms})}} \\
       \cline{3-13}
       & & layers & $d_{model}$ & AUC & recall & nDCG & AUC & recall & nDCG & AUC & recall & nDCG &\\
       \hline
       GRU4Rec & 66338 & 1 & --  & 0.682 & 0.635 & 0.678 & 0.816 & 0.745 & 0.794 & 0.961 & 0.903 & 0.934 & 1.15 \\
       BERT4Rec & 50242 & 1 & 64 & 0.681 & 0.632 & 0.678 & 0.815 & 0.745  & 0.794 & 0.960 & 0.901 & 0.928 & 38.34\\
       BERT4Rec & 55426 & 2 & 32  & 0.721 & 0.665 & 0.698 & 0.815 & 0.745  & 0.794 & 0.960 & 0.902 & 0.941 & 57.53\\
       PinText2 & 69634 & 1 & 256  & 0.558 & 0.541 & 0.608 & 0.716 & 0.669 & 0.691  & 0.883 & 0.817 & 0.782 & 14.46\\
       TiSASRec & 67586 & 2 & 64  & 0.721 & 0.667 & 0.704 & 0.815 & 0.744 & 0.794 & 0.960 & 0.902 & 0.928 & 14.54\\
       ComiRec & 67586 & 1 & 256  & 0.717 & 0.674 & 0.704 & 0.709 & 0.656 & 0.698 & \textbf{0.963} & 0.907 & \textbf{0.979} & 14.61\\
       \hline
       \myname & 49347 & 1 & 32  & \textbf{0.805} & \textbf{0.789} & \textbf{0.781} & \textbf{0.885} & \textbf{0.884} & \textbf{0.909} & 0.930 & \textbf{0.933} & 0.954 & 40.05 \\
       \hline
    \end{tabular}
    }
    \caption{Performance on public datasets. \textit{Params} excludes the parameters in the item embedding table. Recall and nDCG are measured at top-50 items. See Appendix~\ref{ssec:latency_cmpr} for latency measure details.}
    \label{tab:perf_on_cf}
\end{table*}


\textbf{Training strategy.}
We adopt a two-stage setting in the model training in order to enhance the model performance. In the first stage, we fix all the cluster weights to be 1 and train the remaining parameters. After the model converges, we no longer freeze the cluster weights and all parameters are trained until converge. Table \ref{tab:cf_init} summarizes the model performance of two-stage training vs. joint training, and we can clearly see that this strategy is working really well in practice. The rationale behind this is the cluster weights tend to converge too quickly before clusters emerge, therefore we deliberately let the clusters form first before learning the cluster weights. We want to point out this maneuver in our training and hope it can benefit the implementation of similar models. 


\textbf{Results and analysis.}
The performance is summarized in Table~\ref{tab:perf_on_cf}. \myname has a better performance on Amazon and Taobao datasets and is trivially worse than GRU4Rec and ComiRec in AUC on MovieLens. 
Intuitively, the purchase behavior on e-commerce websites (Amazon, Taobao) can be largely explained by the user's interest in multiple categories or brands, while movie-watching is driven more by a movie's popularity and quality rather than the category.
Since all models have very close performance, \myname is still a competitive approach in applications that do not support the strong multi-interest assumption.


\begin{table*}
    \centering
    \resizebox{0.9\textwidth}{!}{
    \begin{tabular}{c|ccc|ccc|ccc}
    \hline\hline 
    \multirow{2}{*}{\makecell{\myname \\ Model}} & \multicolumn{3}{c|}{Amazon} & \multicolumn{3}{c|}{Taobao} & \multicolumn{3}{c}{MovieLens} \\
    \cline{2-10}
    & Epoch & AUC & recall@50 & Epoch & AUC & recall@50 & Epoch & AUC & recall@50 \\
    \hline
    After the first stage & 22 & 0.731 & 0.667 & 4 & 0.821 & 0.749 & 34 & 0.960 & 0.901 \\ 
    After the second stage &  6 & 0.806 & 0.789 & 4 & 0.885 & 0.884 & 6 & 0.930 & 0.933 \\ \hline
    Joint training & 18 & 0.576 & 0.570 &  28 & 0.803 & 0.802 & 14 & 0.924 & 0.937 \\
    \hline
    \end{tabular}
    }
    \caption{Comparison of training strategy and the performance of \myname. The columns \textit{Epoch} shows the training epochs when the best validation AUC is achieved. }
    \label{tab:cf_init}
\end{table*}

\subsection{Using Item Metadata Features}
\label{sec:pin_exp}


\textbf{Dataset.} The dataset contains user engagement history collected from Pinterest, an image-sharing and social media service that allows users to share and discover visual content (images and videos). The interactions between a user and an item (also referred to as a \emph{pin}) are categorized into \textit{impression} (pin is shown to the user), \textit{clickthrough} (user clicks the pin), \textit{re-pin} (user saves the pin into their board collection), and \textit{hide} (user manually hides the pin). 
In total, there are 38 million interactions from 510 thousand users during three weeks of time. Each pin is represented as a 256-dimension feature extracted by the PinSage model \citep{pinsage}. 

User's engagement sequences are processed in a similar way as the public datasets, except that we enforce a one-day gap between the inputs and labels, because adjacent user engagements are usually very similar which makes the prediction task easy.
%
Intuitively, we can use clickthrough and re-pin as the positive label, and hide (which is less often) and impression (without click or re-pin) as the negative label, but since impressions are also recommended to the user at some point, they are likely to be relevant to the user as well, and thus correlate with the positive data. In order to alleviate this bias, we introduce the \textit{random} negative data where pins are sampled from the whole set of pins. The entire negative dataset will consist of 50\% \textit{observed} negative data (hide and impression), and 50\% \textit{random} negative data.

\textbf{Baselines and model configuration.} We compare the multi-interest models PinnerSage and ComiRec, and the single-embedding model TiSASRec with the same setting as in Section~\ref{sec:cf_exp}.

\textbf{Results and analysis.}
As shown in Table~\ref{tab:pin_compare}, \myname outperforms all the state-of-the-art multi-interest sequential models.
\begin{itemize}[noitemsep, topsep=0pt]
    \item PinnerSage shares the same clustering algorithm with \myname, but differs in that 1) each cluster embedding is represented by the medoid of all of its item embeddings; and 2) the cluster weights are heuristic-based (not learned from the model). Instead, \myname learns the cluster representations and weights collectively from data, and thus has a clear advantage over PinnerSage.
    \item TiSASRec has a similar attention module as \myname, except only using the single last attention output as the user embedding. The comparison confirms the necessity of multi-interest representation, as in ComiRec and \myname. 
    \item Compared to ComiRec, \myname interestingly shows that self-attention has stronger representation power than attention with global query. In Appendix~\ref{ssec:syn_exp}, we further use synthetic data to illustrate the fundamental difference between the two types of models. 
\end{itemize}

\begin{table}[t]
    \centering
    \resizebox{\columnwidth}{!}{
        \begin{tabular}{l|cccc}
        \hline\hline
            & precision@20 & recall@20 & AUC & NLL \\
            \hline
            PinnerSage & 0.740 & 0.296 & 0.815 & 1.033\\
            TiSASRec & 0.798 & 0.312 & 0.850 & 0.478 \\
            ComiRec & 0.864 & 0.345 & 0.875 & 0.407 \\
            \hline
            \myname & \textbf{0.882} & \textbf{0.353} & \textbf{0.893} & \textbf{0.377} \\
            \hline
        \end{tabular}
    }
        \caption{Performance on the Pinterest dataset.}
        \label{tab:pin_compare}
\end{table}

\begin{table}
    \centering
    \resizebox{\columnwidth}{!}{
    \begin{tabular}{cc|cc|cc|cc}
    \hline\hline
     \multicolumn{2}{c|}{NLL} & \multicolumn{2}{c|}{\makecell{Triplet \\ ($\alpha=$0.2)}} & \multicolumn{2}{c|}{\makecell{Triplet \\ ($\alpha=$0.5)}} & \multicolumn{2}{c}{\makecell{Triplet \\ ($\alpha=$0.8)}}\\
     \hline
      AUC & R@50 & AUC & R@50 & AUC & R@50 & AUC & R@50 \\
      \hline
      0.930 & 0.933 & 0.885 & 0.882 & 0.903 & 0.901 & 0.906 & 0.906 \\
      \hline
    \end{tabular}
    }
    \caption{\myname on MovieLens with different loss functions.}
    \label{tab:loss_func_cmpr}
\end{table}


\subsection{Ablation Study}
\label{sec:ablation_study}

This section focuses on the claim that 1) the user's preference should be dynamically learned from the temporal pattern by demonstrating the effectiveness of the learned personalized cluster weights module of \myname; 2) The NLL loss is sufficient to reach good performance without the need to sweep the hyperparameters required in other losses, \emph{e.g.} triplet loss; 3) effectiveness of positional and temporal encoding (Section~\ref{ssec:variants_positional_temporal_encoding}). We refer the readers to Appendix for a comprehensive ablation study on attention mechanism (\ref{ssec:syn_exp}), re-configuration and comparison of the clustering method post-training (\ref{ssec:abl_cluster}). We will focus on the ablation study on interest weights and loss function in this section.


\begin{figure*}
\centering
\begin{minipage}{0.56\textwidth}
    \centering
    \includegraphics[height=1.65in]{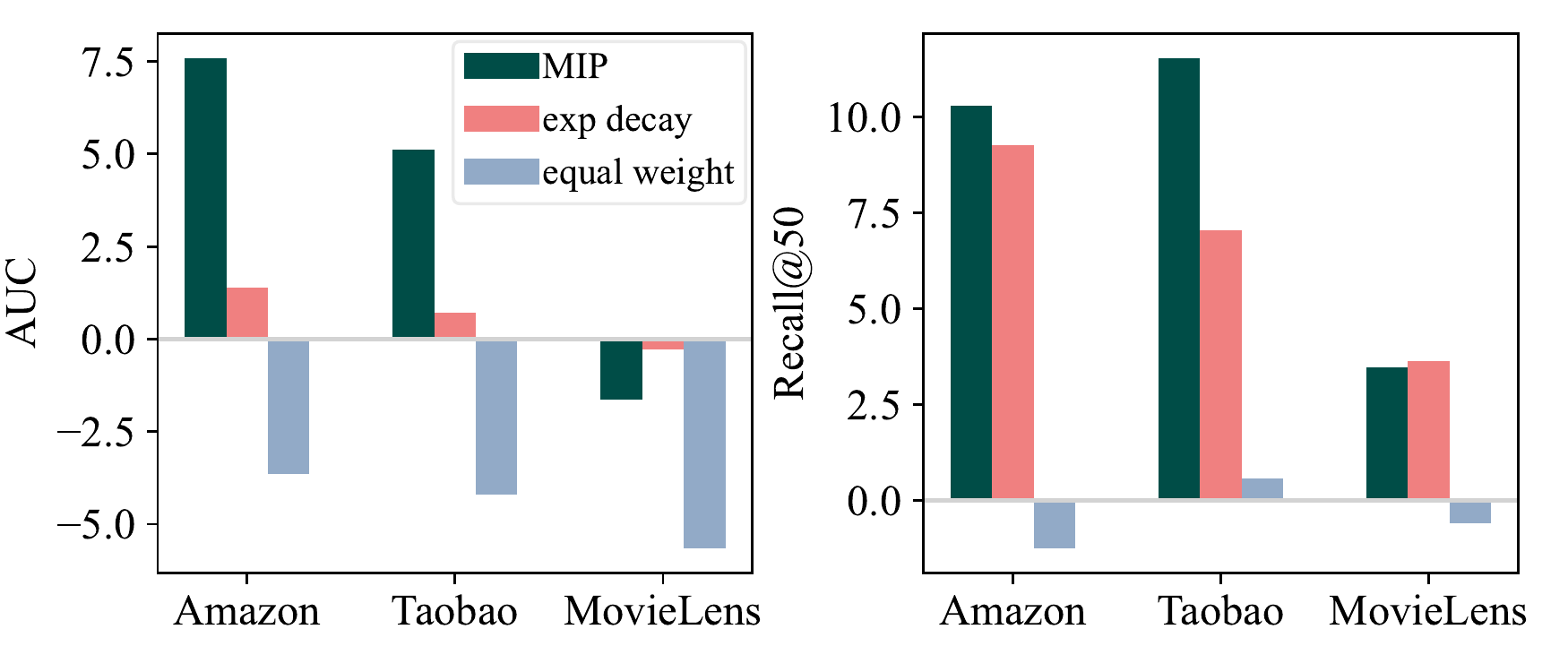}
    \\ (a) Public datasets
\end{minipage}
\begin{minipage}{0.32\textwidth}
\centering
\includegraphics[height=1.65in]{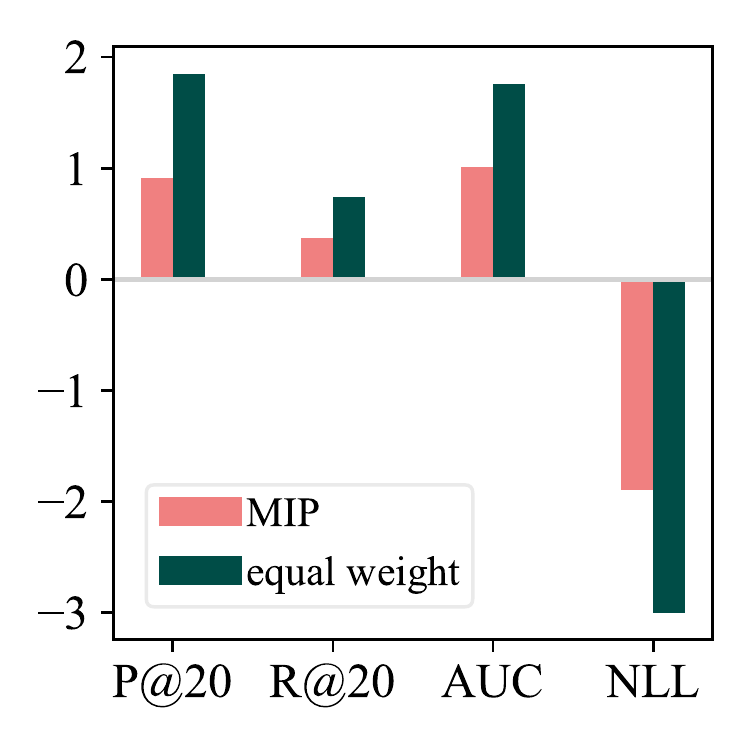}
    \\ (b) Pinterest dataset
\end{minipage}
\caption{Cluster weight variants performance (difference to the best baseline models, in the unit of $10^{-2}$)}
\label{fig:abl_weight}
\end{figure*}

\textbf{Interest weights.}\quad To validate the assumption that the preference trends (weights of multiple interests) change from user to user, we compare the \myname with two variants that disable the cluster weight module. The first one (referred to as \textit{\myname(Equal Weight)}) constantly assigns 1 to the cluster weights, leading to an equally weighted multi-interest user representation. Another one (referred to as \textit{\myname(Exp Decay)}) uses an exponential-decay heuristic weights given by $w_\lambda=\sum_{i: C_i \in \lambda}exp(-\epsilon(t_{now} - t_{i})) $\citep{pinnersage}. We let $t_{now}$ be the last user engagement time $t_{last\_item}$, since in practice it's unrealistic to update the weights in real-time. According to \citet{pinnersage}, we also set $\epsilon=0.01$, which balances well between emphasizing recently engaged items and accentuating frequently engaged categories. We let the number of clusters be $\Lambda=5$. 
%
We calculate performance on these model variants and show the relative percentage difference from the best baseline model in Figure~\ref{fig:abl_weight}. Note that smaller NLL means better performance. In all experiments, the AUC and recall can benefit from learning cluster weights (\myname). The second best variant is usually having an equal weight for all clusters. 


\textbf{Loss function.}\quad Beside the NLL loss function in Equation~\ref{eq:nll}, triplet loss is also widely used in  recommendation system and contrastive learning. Let the $y^+$ be the similarity prediction between the user representation and a positive item, and $y^-$ be the predicted similarity between user representation and a negative item. The triplet loss is given by:



\begin{equation}
\small
\mathcal{L}_{\alpha} = \sum(y^+ - y^- + \alpha)_{\mu \in \mathcal{U}} 
\end{equation}

where $\alpha$ is a hyperparameter of the positive-negative margin. With triplet loss, we added a linear factor $\beta$ (\textit{learned}) in Equation~\ref{eq:aggregator}, (\emph{i.e.} $y=max\{\beta w_\lambda \cdot (\bm{z}_\lambda \cdot \bm{p})\}^\Lambda_{\lambda=1}$) to re-scale the similarity since $y$ is unbounded and makes the choice of $\alpha$ to be hard. We let the \myname train on the MovieLens dataset to investigate the impact on the performance of loss function choice. The results in Table~\ref{tab:loss_func_cmpr} illustrate that the triplet loss marginally underperforms the NLL loss we used.

\subsection{Positional and Temporal Encoding}
\label{ssec:variants_positional_temporal_encoding}

In Equation~\ref{eq:concat_encoding}, the sequential (positional) and temporal information are encoded and included in the self-attention module to produce the multi-interest representations. The motivation is that given items from the same category, the recent ones might better represent the user's current interest than the obsolete items. We verify 1) if the incorporation of positional and temporal encoding is critical to the performance; 2) how the encoding (Equation~\ref{eq:encoding}) method affects the performance. 

\textbf{Configuration.} The \myname are configured on the two set of choices: the Equation~\ref{eq:concat_encoding} can be configured alternatively: 
\begin{itemize}[noitemsep, topsep=0pt]
 \item item embedding only: $\bm{e}_j=\bm{p}_j$.
 \item + positional: $\bm{e}_j = [\bm{p}_j;\bm{\rho}(j)]$, where $\bm{\rho}(j)$ is given by Equation~\ref{eq:encoding}.
 \item + temporal: $\bm{e}_j = [\bm{p}_j; \bm{\tau}(t_j)]$, where $\bm{\tau}(t_j)$ is given by Equation~\ref{eq:encoding}.
 \item + positional and temporal: Equation~\ref{eq:concat_encoding} and Equation~\ref{eq:encoding}
\end{itemize}
and there are several other choices of temporal encodings other than the sinusoidal form in the Equation~\ref{eq:encoding}:
\begin{itemize}[noitemsep, topsep=0pt]
    \item One-hot \citep{zhou2018atrank}: Create exponential buckets $[0, b)$, $[b, b^2)$, $\cdots$, $[b^{k-1}, \infty)$ with base $b$, and encode the timestamp as an one-hot vector, \emph{i.e.} $\tau_i = lookup(buckets(t))$.
    \item Two-hot \citep{ccnn}: Similarly create exponential boundaries $\{0$, $b$, $b^2$, $\cdots$, $b^{k-1}$, $\infty\}$, and encode the timestamp as $\mathbf{\bm{\tau}}_i = log_b(t) - i$ and $\mathbf{\bm{\tau}}_{i + 1} = i + 1 - log_b(t)$, where $b^i \leq t < b^{i+1}$.
\end{itemize}

\textbf{Dataset and Results.} The options to include positional and temporal information are evaluated on all the dataset (Table~\ref{tab:hps_ablation_cf}), and the encodings methods are compared exhaustively on Pinterest dataset (Table~\ref{tab:tem_pos}). 

\begin{table}
    \begin{tabular}{c|c|cc}
    \hline\hline
        Dataset & Configuration & AUC & Recall \\
        \hline
        \multirow{4}{*}{Amazon} & item embedding &  76.34 & 73.75 \\
        & +positional & 76.19 & 74.28 \\
        & +temporal & 78.59 & 76.94 \\
        & +both & \textbf{79.31} & \textbf{78.22} \\
    \hline        
    \multirow{4}{*}{Taobao} & item embedding  & 82.06 & 81.75 \\
        & +positional & 86.54 & \textbf{86.22} \\
        & +temporal & \textbf{86.72} & 86.56  \\
        & +both & 86.59 & 85.83 \\
    \hline
    \multirow{4}{*}{MovieLens} & item embedding  & \textbf{95.26} & \textbf{94.45}\\
        & +positional & 95.01 & 94.31\\
        & +temporal & 94.96 & 94.19  \\
        & + both & 94.61 & 94.12  \\
    \hline
    \end{tabular}
    \caption{Ablation study on the positional and temporal encoding on public datasets. (in $10^{-2}$)}
    \label{tab:hps_ablation_cf}
\end{table}

\textbf{Analysis.} Two conclusions can be made from Table~\ref{tab:hps_ablation_cf} and Table~\ref{tab:tem_pos}: 1) including both temporal and positional information is a safe option, which has the best performance on Amazon and Pinterest and marginally ($<0.01$) worse performance on Taobao and MovieLens; and 2) the model is insensitive to encoding methods. 




\begin{table}[t]
    \centering
    \begin{tabular}{l|cc}
    \hline\hline
    Configuration & AUC & NLL \\
    \hline
      item embedding   & 0.8923 & \textbf{0.377} \\
      + positional   &  0.8846 & 0.386\\
      + temporal (one-hot) &  0.8850 & 0.388\\
      + temporal (two-hot) &  0.8846 & 0.385 \\
      + temporal (sinusoid) &  0.8921 & \textbf{0.377}\\
      + both (one-hot) &  0.8861 & 0.382 \\
      + both (two-hot) &  0.8852 & 0.387\\
      + both (sinusoid) &  \textbf{0.8926} & \textbf{0.377}\\
    \hline
    \end{tabular}
    \caption{Influence of temporal and positional encoding in attention on the performance in \myname}
    \label{tab:tem_pos}
\end{table}

\label{sec:experiment}

\section{Conclusions}

In this paper, we study the problem of multi-interest user embedding for recommendation systems. 
We follow the recent findings on representing users with multiple embeddings, which has been proven helpful over the single user representation. In addition, we illustrate that in industrial recommendation systems, it is important to have a set of weights for these multiple embeddings for a more efficient candidate generation process due to its budget on the number of items returned.
More specifically, we define the likelihood of an engagement based on the \textit{closest} user embedding to the item embedding and update the weight for the corresponding cluster. Moreover, the case studies on multiple real-world datasets have demonstrated our advantage over state-of-the-art approaches. Finally, the ablation study on the cluster weight module demonstrated our intuition that simple heuristic does not work as well as personalized model-learned interest weights. We refer readers to the appendix for extensive study on other model design choices. 

\label{sec:conclusion}


\nocite{langley00}

\bibliography{ref}

\begin{thebibliography}{41}
\providecommand{\natexlab}[1]{#1}
\providecommand{\url}[1]{\texttt{#1}}
\expandafter\ifx\csname urlstyle\endcsname\relax
  \providecommand{\doi}[1]{doi: #1}\else
  \providecommand{\doi}{doi: \begingroup \urlstyle{rm}\Url}\fi

\bibitem[Cen et~al.(2020)Cen, Zhang, Zou, Zhou, Yang, and Tang]{comirec}
Cen, Y., Zhang, J., Zou, X., Zhou, C., Yang, H., and Tang, J.
\newblock Controllable multi-interest framework for recommendation.
\newblock In \emph{Proceedings of the 26th ACM SIGKDD International Conference
  on Knowledge Discovery \& Data Mining}, pp.\  2942--2951, 2020.

\bibitem[Chen et~al.(2019)Chen, Zhang, and Qin]{chen2019dynamic}
Chen, X., Zhang, Y., and Qin, Z.
\newblock Dynamic explainable recommendation based on neural attentive models.
\newblock In \emph{Proceedings of the AAAI Conference on Artificial
  Intelligence}, volume~33, pp.\  53--60, 2019.

\bibitem[Cheng et~al.(2016)Cheng, Koc, Harmsen, Shaked, Chandra, Aradhye,
  Anderson, Corrado, Chai, Ispir, et~al.]{cheng2016wide}
Cheng, H.-T., Koc, L., Harmsen, J., Shaked, T., Chandra, T., Aradhye, H.,
  Anderson, G., Corrado, G., Chai, W., Ispir, M., et~al.
\newblock Wide \& deep learning for recommender systems.
\newblock In \emph{Proceedings of the 1st workshop on deep learning for
  recommender systems}, pp.\  7--10, 2016.

\bibitem[Covington et~al.(2016)Covington, Adkan, and Sargin]{covington2016deep}
Covington, P., Adkan, J., and Sargin, E.
\newblock Deep neural networks for youtube recommendations.
\newblock In \emph{Proceedings of the 10th ACM conference on recommender
  systems}, pp.\  191--198, 2016.

\bibitem[Devooght \& Bersini(2017)Devooght and Bersini]{devooght2017long}
Devooght, R. and Bersini, H.
\newblock Long and short-term recommendations with recurrent neural networks.
\newblock In \emph{Proceedings of the 25th Conference on User Modeling,
  Adaptation and Personalization}, pp.\  13--21, 2017.

\bibitem[Donkers et~al.(2017)Donkers, Loepp, and
  Ziegler]{donkers2017sequential}
Donkers, T., Loepp, B., and Ziegler, J.
\newblock Sequential user-based recurrent neural network recommendations.
\newblock In \emph{Proceedings of the Eleventh ACM Conference on Recommender
  Systems}, pp.\  152--160, 2017.

\bibitem[Epasto \& Perozzi(2019)Epasto and Perozzi]{epasto2019single}
Epasto, A. and Perozzi, B.
\newblock Is a single embedding enough? learning node representations that
  capture multiple social contexts.
\newblock In \emph{The World Wide Web Conference}, pp.\  394--404, 2019.

\bibitem[Ester et~al.(1996)Ester, Kriegel, Sander, Xu, et~al.]{dbscan}
Ester, M., Kriegel, H.-P., Sander, J., Xu, X., et~al.
\newblock A density-based algorithm for discovering clusters in large spatial
  databases with noise.
\newblock In \emph{kdd}, volume~96, pp.\  226--231, 1996.

\bibitem[He \& McAuley(2016)He and McAuley]{he2016fusing}
He, R. and McAuley, J.
\newblock Fusing similarity models with markov chains for sparse sequential
  recommendation.
\newblock In \emph{2016 IEEE 16th International Conference on Data Mining
  (ICDM)}, pp.\  191--200. IEEE, 2016.

\bibitem[Hidasi et~al.(2015)Hidasi, Karatzoglou, Baltrunas, and Tikk]{gru4rec}
Hidasi, B., Karatzoglou, A., Baltrunas, L., and Tikk, D.
\newblock Session-based recommendations with recurrent neural networks.
\newblock \emph{arXiv preprint arXiv:1511.06939}, 2015.

\bibitem[Jiang et~al.(2020)Jiang, Wang, Wei, Gao, Wang, and
  Nie]{jiang2020aspect}
Jiang, H., Wang, W., Wei, Y., Gao, Z., Wang, Y., and Nie, L.
\newblock What aspect do you like: Multi-scale time-aware user interest
  modeling for micro-video recommendation.
\newblock In \emph{Proceedings of the 28th ACM International Conference on
  Multimedia}, pp.\  3487--3495, 2020.

\bibitem[Kang \& McAuley(2018)Kang and McAuley]{kang2018self}
Kang, W.-C. and McAuley, J.
\newblock Self-attentive sequential recommendation.
\newblock In \emph{2018 IEEE International Conference on Data Mining (ICDM)},
  pp.\  197--206. IEEE, 2018.

\bibitem[Li et~al.(2019)Li, Liu, Wu, Xu, Zhao, Huang, Kang, Chen, Li, and
  Lee]{li2019multi}
Li, C., Liu, Z., Wu, M., Xu, Y., Zhao, H., Huang, P., Kang, G., Chen, Q., Li,
  W., and Lee, D.~L.
\newblock Multi-interest network with dynamic routing for recommendation at
  tmall.
\newblock In \emph{Proceedings of the 28th ACM International Conference on
  Information and Knowledge Management}, pp.\  2615--2623, 2019.

\bibitem[Li et~al.(2020)Li, Wang, and McAuley]{tisas}
Li, J., Wang, Y., and McAuley, J.
\newblock Time interval aware self-attention for sequential recommendation.
\newblock In \emph{Proceedings of the 13th International Conference on Web
  Search and Data Mining}, pp.\  322--330, 2020.

\bibitem[Li et~al.(2005)Li, Lu, and Xuefeng]{li2005hybrid}
Li, Y., Lu, L., and Xuefeng, L.
\newblock A hybrid collaborative filtering method for multiple-interests and
  multiple-content recommendation in e-commerce.
\newblock \emph{Expert systems with applications}, 28\penalty0 (1):\penalty0
  67--77, 2005.

\bibitem[Pal et~al.(2020)Pal, Eksombatchai, Zhou, Zhao, Rosenberg, and
  Leskovec]{pinnersage}
Pal, A., Eksombatchai, C., Zhou, Y., Zhao, B., Rosenberg, C., and Leskovec, J.
\newblock Pinnersage: Multi-modal user embedding framework for recommendations
  at pinterest.
\newblock In \emph{Proceedings of the 26th ACM SIGKDD International Conference
  on Knowledge Discovery \& Data Mining}, pp.\  2311--2320, 2020.

\bibitem[P{\'e}rez et~al.(2019)P{\'e}rez, Marinkovi{\'c}, and
  Barcel{\'o}]{perez2019turing}
P{\'e}rez, J., Marinkovi{\'c}, J., and Barcel{\'o}, P.
\newblock On the turing completeness of modern neural network architectures.
\newblock \emph{arXiv preprint arXiv:1901.03429}, 2019.

\bibitem[Quadrana et~al.(2017)Quadrana, Karatzoglou, Hidasi, and
  Cremonesi]{quadrana2017personalizing}
Quadrana, M., Karatzoglou, A., Hidasi, B., and Cremonesi, P.
\newblock Personalizing session-based recommendations with hierarchical
  recurrent neural networks.
\newblock In \emph{proceedings of the Eleventh ACM Conference on Recommender
  Systems}, pp.\  130--137, 2017.

\bibitem[Rendle et~al.(2010)Rendle, Freudenthaler, and
  Schmidt-Thieme]{rendle2010factorizing}
Rendle, S., Freudenthaler, C., and Schmidt-Thieme, L.
\newblock Factorizing personalized markov chains for next-basket
  recommendation.
\newblock In \emph{Proceedings of the 19th international conference on World
  wide web}, pp.\  811--820, 2010.

\bibitem[Sabour et~al.(2017)Sabour, Frosst, and Hinton]{sabour2017dynamic}
Sabour, S., Frosst, N., and Hinton, G.~E.
\newblock Dynamic routing between capsules.
\newblock \emph{arXiv preprint arXiv:1710.09829}, 2017.

\bibitem[Shi et~al.(2021)Shi, Zhang, Wu, Chang, Qian, Hasegawa-Johnson, and
  Zhao]{ccnn}
Shi, H., Zhang, Y., Wu, H., Chang, S., Qian, K., Hasegawa-Johnson, M., and
  Zhao, J.
\newblock Continuous cnn for nonuniform time series.
\newblock In \emph{ICASSP 2021-2021 IEEE International Conference on Acoustics,
  Speech and Signal Processing (ICASSP)}, pp.\  3550--3554. IEEE, 2021.

\bibitem[Shi et~al.(2022)Shi, Gao, Tian, Chen, and Zhao]{shi2022learning}
Shi, H., Gao, S., Tian, Y., Chen, X., and Zhao, J.
\newblock Learning bounded context-free-grammar via lstm and the transformer:
  Difference and the explanations.
\newblock In \emph{Proceedings of the AAAI Conference on Artificial
  Intelligence}, volume~36, pp.\  8267--8276, 2022.

\bibitem[Shi \& Malik(2000)Shi and Malik]{spectral_clustering}
Shi, J. and Malik, J.
\newblock Normalized cuts and image segmentation.
\newblock \emph{IEEE Transactions on pattern analysis and machine
  intelligence}, 22\penalty0 (8):\penalty0 888--905, 2000.

\bibitem[Sun et~al.(2019)Sun, Liu, Wu, Pei, Lin, Ou, and
  Jiang]{sun2019bert4rec}
Sun, F., Liu, J., Wu, J., Pei, C., Lin, X., Ou, W., and Jiang, P.
\newblock Bert4rec: Sequential recommendation with bidirectional encoder
  representations from transformer.
\newblock In \emph{Proceedings of the 28th ACM international conference on
  information and knowledge management}, pp.\  1441--1450, 2019.

\bibitem[Vaswani et~al.(2017)Vaswani, Shazeer, Parmar, Uszkoreit, Jones, Gomez,
  Kaiser, and Polosukhin]{vaswani2017attention}
Vaswani, A., Shazeer, N., Parmar, N., Uszkoreit, J., Jones, L., Gomez, A.~N.,
  Kaiser, L., and Polosukhin, I.
\newblock Attention is all you need.
\newblock \emph{arXiv preprint arXiv:1706.03762}, 2017.

\bibitem[Wandabwa et~al.(2020)Wandabwa, Naeem, Mirza, Pears, and
  Nguyen]{wandabwa2020multi}
Wandabwa, H., Naeem, M.~A., Mirza, F., Pears, R., and Nguyen, A.
\newblock Multi-interest user profiling in short text microblogs.
\newblock In \emph{International Conference on Design Science Research in
  Information Systems and Technology}, pp.\  154--168. Springer, 2020.

\bibitem[Wang(2007)]{wang2007multi}
Wang, F.
\newblock Multi-interest communities and community-based recommendation.
\newblock 2007.

\bibitem[Wang et~al.(2018)Wang, Huang, Zhao, Zhang, Zhao, and
  Lee]{wang2018billion}
Wang, J., Huang, P., Zhao, H., Zhang, Z., Zhao, B., and Lee, D.~L.
\newblock Billion-scale commodity embedding for e-commerce recommendation in
  alibaba.
\newblock In \emph{Proceedings of the 24th ACM SIGKDD International Conference
  on Knowledge Discovery \& Data Mining}, pp.\  839--848, 2018.

\bibitem[Wang et~al.(2019)Wang, He, Wang, Feng, and Chua]{wang2019neural}
Wang, X., He, X., Wang, M., Feng, F., and Chua, T.-S.
\newblock Neural graph collaborative filtering.
\newblock In \emph{Proceedings of the 42nd international ACM SIGIR conference
  on Research and development in Information Retrieval}, pp.\  165--174, 2019.

\bibitem[Ward~Jr(1963)]{ward1963hierarchical}
Ward~Jr, J.~H.
\newblock Hierarchical grouping to optimize an objective function.
\newblock \emph{Journal of the American statistical association}, 58\penalty0
  (301):\penalty0 236--244, 1963.

\bibitem[Weston et~al.(2013)Weston, Weiss, and Yee]{weston2013nonlinear}
Weston, J., Weiss, R.~J., and Yee, H.
\newblock Nonlinear latent factorization by embedding multiple user interests.
\newblock In \emph{Proceedings of the 7th ACM conference on Recommender
  systems}, pp.\  65--68, 2013.

\bibitem[Xia et~al.(2017)Xia, Li, Li, and Li]{xia2017attention}
Xia, B., Li, Y., Li, Q., and Li, T.
\newblock Attention-based recurrent neural network for location recommendation.
\newblock In \emph{2017 12th International Conference on Intelligent Systems
  and Knowledge Engineering (ISKE)}, pp.\  1--6. IEEE, 2017.

\bibitem[Xu et~al.(2019)Xu, Zhao, Liu, Xu, S.~Sheng, Cui, Zhou, and
  Xiong]{xu2019recurrent}
Xu, C., Zhao, P., Liu, Y., Xu, J., S.~Sheng, V. S.~S., Cui, Z., Zhou, X., and
  Xiong, H.
\newblock Recurrent convolutional neural network for sequential recommendation.
\newblock In \emph{The World Wide Web Conference}, pp.\  3398--3404, 2019.

\bibitem[Xue et~al.(2005)Xue, Lin, Yang, Xi, Zeng, Yu, and
  Chen]{xue2005scalable}
Xue, G.-R., Lin, C., Yang, Q., Xi, W., Zeng, H.-J., Yu, Y., and Chen, Z.
\newblock Scalable collaborative filtering using cluster-based smoothing.
\newblock In \emph{Proceedings of the 28th annual international ACM SIGIR
  conference on Research and development in information retrieval}, pp.\
  114--121, 2005.

\bibitem[Ying et~al.(2018)Ying, He, Chen, Eksombatchai, Hamilton, and
  Leskovec]{pinsage}
Ying, R., He, R., Chen, K., Eksombatchai, P., Hamilton, W.~L., and Leskovec, J.
\newblock Graph convolutional neural networks for web-scale recommender
  systems.
\newblock In \emph{Proceedings of the 24th ACM SIGKDD International Conference
  on Knowledge Discovery \& Data Mining}, pp.\  974--983, 2018.

\bibitem[You et~al.(2019)You, Wang, Pal, Eksombatchai, Rosenburg, and
  Leskovec]{you2019hierarchical}
You, J., Wang, Y., Pal, A., Eksombatchai, P., Rosenburg, C., and Leskovec, J.
\newblock Hierarchical temporal convolutional networks for dynamic recommender
  systems.
\newblock In \emph{The world wide web conference}, pp.\  2236--2246, 2019.

\bibitem[Yu(2008)]{yu2008using}
Yu, L.
\newblock Using ontology to enhance collaborative recommendation based on
  community.
\newblock In \emph{2008 The Ninth International Conference on Web-Age
  Information Management}, pp.\  45--49. IEEE, 2008.

\bibitem[Yue \& Xiang(2012)Yue and Xiang]{yue2012multi}
Yue, W. and Xiang, C.
\newblock A multi-interests model of recommendation system based on customer
  life cycle.
\newblock In \emph{2012 Fifth International Conference on Intelligent
  Computation Technology and Automation}, pp.\  22--25. IEEE, 2012.

\bibitem[Zhang et~al.(1996)Zhang, Ramakrishnan, and Livny]{zhang1996birch}
Zhang, T., Ramakrishnan, R., and Livny, M.
\newblock Birch: an efficient data clustering method for very large databases.
\newblock \emph{ACM sigmod record}, 25\penalty0 (2):\penalty0 103--114, 1996.

\bibitem[Zhou et~al.(2018)Zhou, Bai, Song, Liu, Zhao, Chen, and
  Gao]{zhou2018atrank}
Zhou, C., Bai, J., Song, J., Liu, X., Zhao, Z., Chen, X., and Gao, J.
\newblock Atrank: An attention-based user behavior modeling framework for
  recommendation.
\newblock In \emph{Proceedings of the AAAI Conference on Artificial
  Intelligence}, volume~32, 2018.

\bibitem[Zhuang et~al.(2020)Zhuang, Zhao, Subramanian, Lin, Krishnapuram, and
  Zwol]{pintext2}
Zhuang, J., Zhao, J., Subramanian, Srinivas, A., Lin, Y., Krishnapuram, B., and
  Zwol, R.~v.
\newblock Pintext 2: Attentive bag of annotations embedding.
\newblock In \emph{Proceedings of DLP-KDD 2020}, 2020.

\end{thebibliography}
\bibliographystyle{icml2023}

\newpage
\appendix
\onecolumn
\section{Appendix}

\subsection{Variants of Attention Mechanism}
\begin{table*}
    \centering
    \begin{tabular}{l|c|c|c}
    \hline\hline
     \multirow{2}{*}{Model}  & \multicolumn{2}{c|}{Global query} & \multirow{2}{*}{Self-attention} \\
     \cline{2-3}
     & Non-shared  &  Shared \\
     \hline
     Key vector $\bm{k}_j$ & \multicolumn{3}{c}{$(W^h_r \bm{p}_j + \bm{b}^h)$} \\
     \hline
     Query vector & $\bm{q}^h$ & $\bm{q}$ & $\bm{q}_i^h = W^h_q \bm{p}_i + \bm{b}^h_q $\\
     \hline
     $\bm{q}$ shared between sequences & \multicolumn{2}{c|}{Yes} & No \\
     \hline
     $\bm{q}$ shared between att. heads & No & Yes &  No \\
     \hline
     Dot-product    & $e^h_j = \bm{q}^{h\top} \cdot \bm{k}_j^h $ &
     $e^h_j = \bm{q}^\top \cdot \bm{k}_j^h$ &
    $e^h_{i, j} = \bm{q}_i^{h\top} \cdot \bm{k}_j^h$ \\
    \hline
    \end{tabular}
    \caption{Variations of multi-head attention. 
    }
    \label{tab:att_models}
\end{table*}


To interpret the performance improvement of our models against other attention models that have been applied in the recommendation system, we further construct a synthetic dataset and visualize the internal attention and the user representations aggregated from different attentions. 

\textbf{Synthetic dataset. } Without loss of generality, we assume there are $G$ global clusters in the corpus, representing different global categories, each of which is a $d$-dimensional Gaussian distribution. Each user is interested in up to $k$ ($<G$) categories, referred to as user clusters. We generate the oracle user interest model by sampling no more than $k$ clusters from $G$ global clusters following a multinomial distribution. Then each of the items in the user engagement history is sampled in two steps: uniformly sample one cluster from user clusters, then sample from the $d$-dimensional Gaussian distribution. Note that in the synthetic model, the item-to-cluster affinity is measured in Euclidean distance, while in the recommendation model, the affinity is decided by cosine distance. To eliminate this discrepancy, we force the Gaussian distributions to center on the unit sphere, so that the rankings by cosine distance and Euclidean distance are consistent. 

We use a 2D dataset ($d=2$) for visualization purposes and another high-dimensional dataset for quantitative evaluation. For the 2D dataset, we set a relatively small $G=8$ and $k=4$ in order to have a clear boundary between clusters. Since there are only 162 distinct subsets\footnote{Number of ways to select no more than 4 clusters from a pool of 8 clusters: $162 = \binom{8}{1} + \binom{8}{2} + \binom{8}{3} + \binom{8}{4}$}
with $G=8, k=4$, we use 100 of them for training, 31 for validation, and the remaining 31 for testing. For the high-dimensional dataset, we set $G=1024$ and $k=8$, and let $d=16,32,64,128$. We generate 10000 users for training, 1000 for validation, and another 1000 for testing.

\begin{figure}[pt]
\centering
\begin{minipage}{0.48\columnwidth}
    \centering
    \includegraphics[width=\columnwidth]{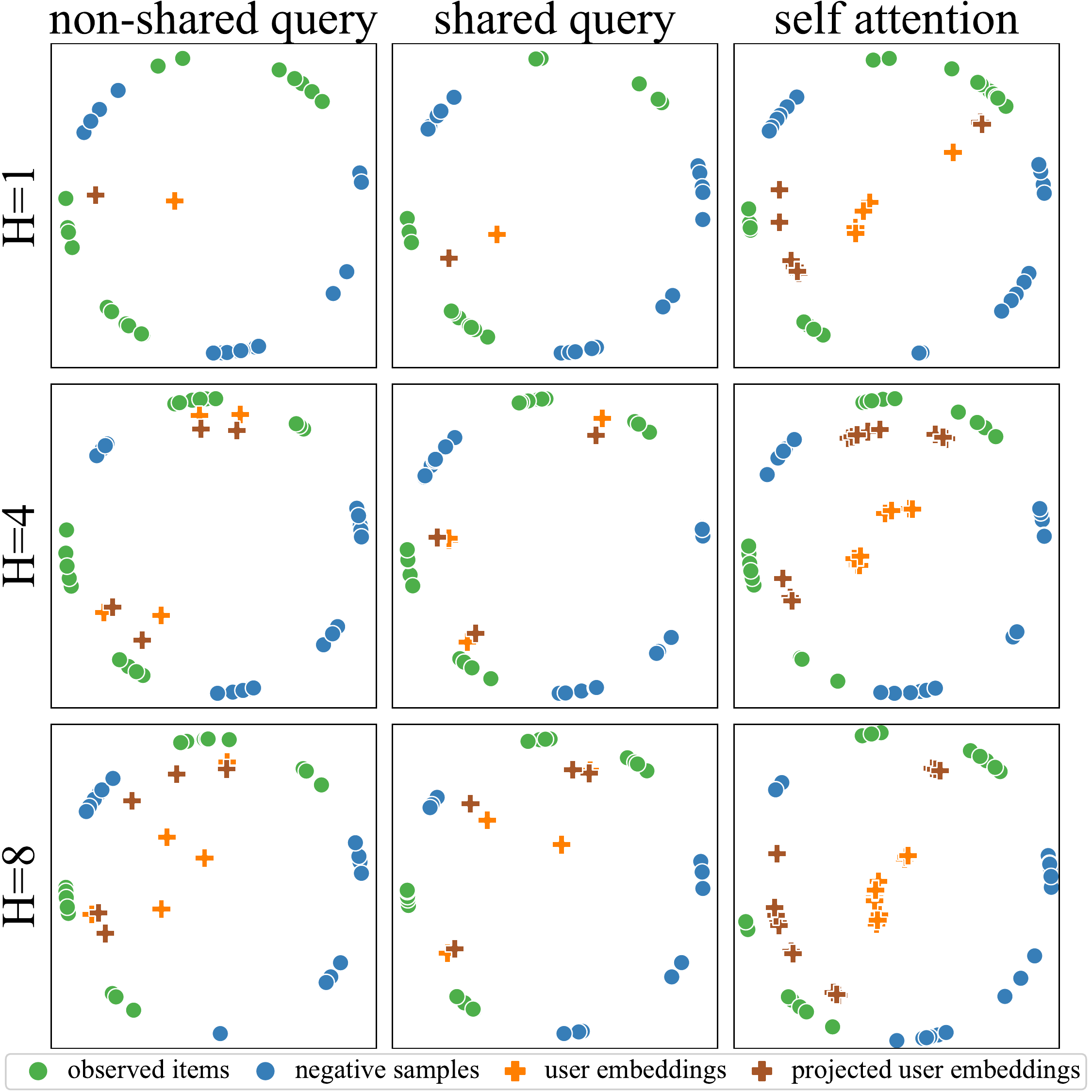}
    \caption{Learned user representations with different attention mechanisms. 
    Non-shared and shared global query results might miss some user interest or are close to the negative categories, while self-attention results are comprehensive and accurate. }
    \label{fig:user_vis}
\end{minipage}
\hspace{0.1in}
\begin{minipage}{0.48\columnwidth}
    \centering
    \includegraphics[width=2.9in]{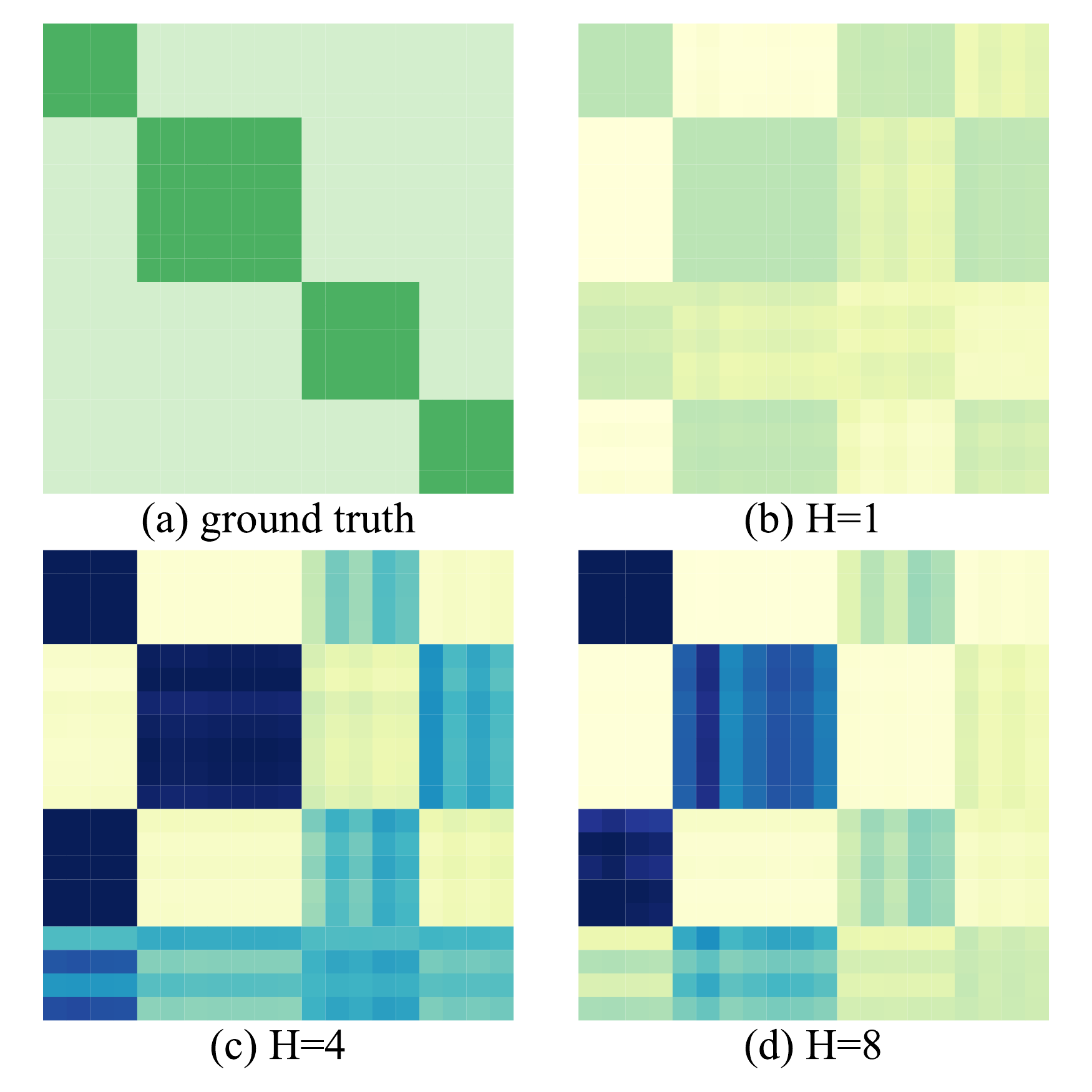}
    \caption{Learned attention scores in self-attention model. Darker color represents higher values. (a) the indicator function $\mathbbm{1}_{[\mathcal{C}_i = \mathcal{C}_j]}$, (b-d) $a_{i, j}$. The input sequence is re-ordered for better visualization. }
    \label{fig:self_att_vis}
\end{minipage}
\end{figure}

\textbf{Attention models. } We focus on comparing our attention model (i.e. \textit{self-attention}), the attention model utilized in ComiRec (i.e. \textit{Non-shared query}), and the one used in PinText2 (i.e. \textit{Shared query}). The comparison of the attentions is in Table~\ref{tab:att_models}.

%
%
%
%
For simplicity, we remove the temporal and positional encoding from the computation of attentions, skip the Ward clustering step from \myname, and directly represent user as Equation~\ref{eq:z}.
Also, the dropout layer is removed in order to eliminate randomness in visualization.

\textbf{Metrics.} We visualize the intermediate results and user representations learned from the 2D dataset for qualitative evaluation. For high-dimensional data, we evaluate the performance by AUC and normalized discounted cumulative gain (nDCG). 

\begin{figure}[t]
\centering
\begin{minipage}{0.49\columnwidth}
    \centering
    \includegraphics[width=\columnwidth]{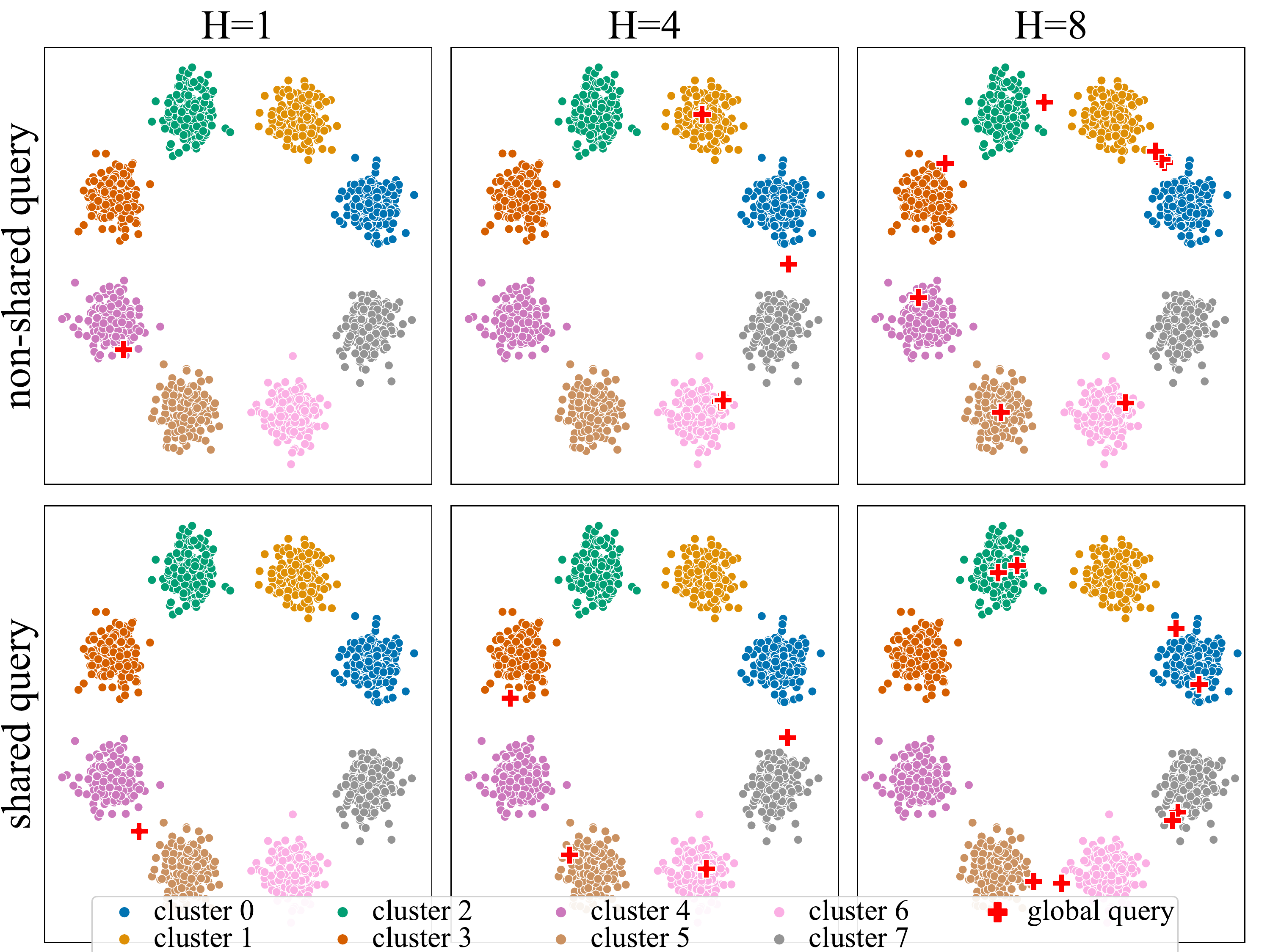}
    \caption{Learned global interests in global query models. The query vector is reversely projected and normalized.}
    \label{fig:global_query}
\end{minipage}
%
\begin{minipage}{0.49\columnwidth}
    \centering
    \includegraphics[width=\columnwidth]{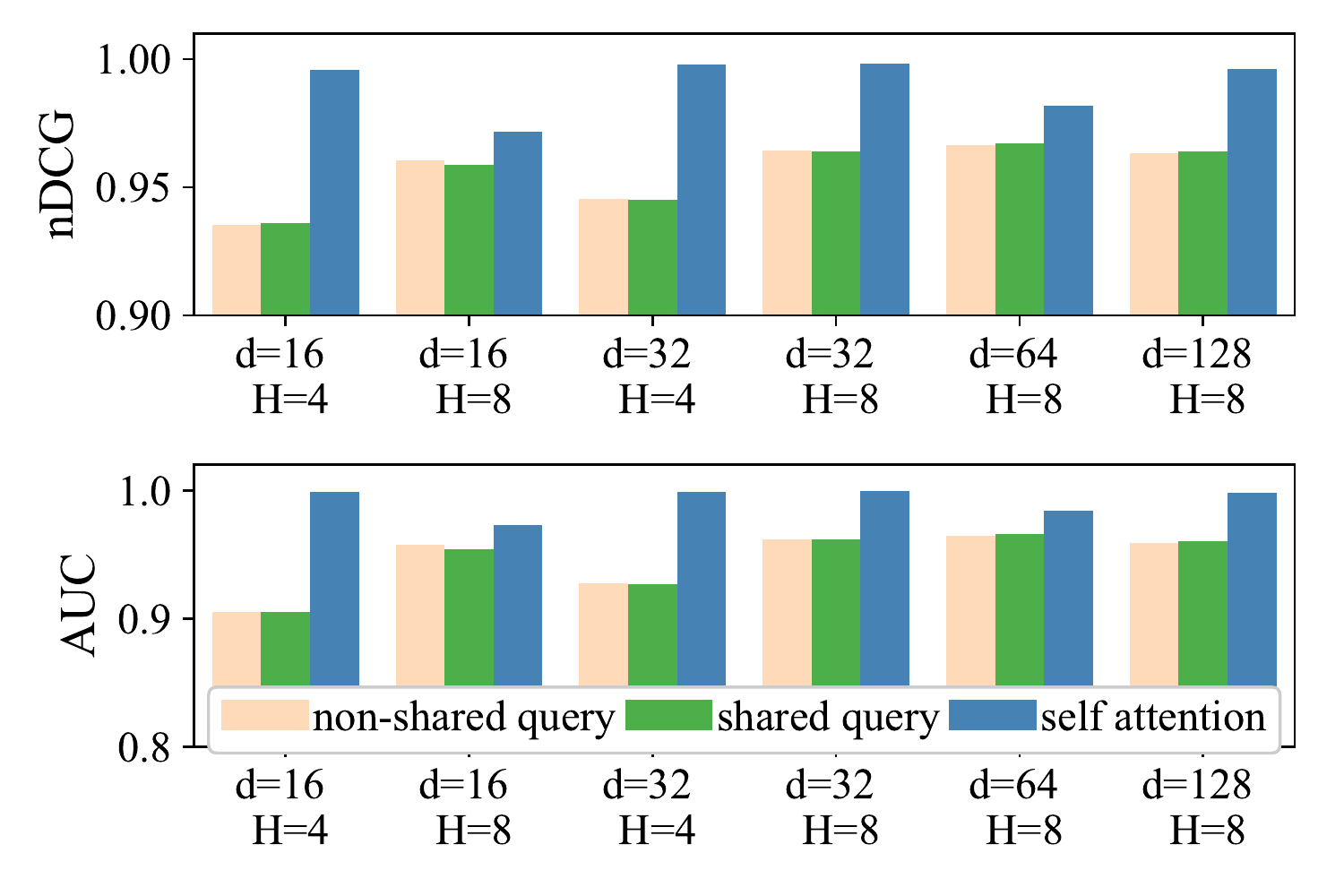}
    \caption{Performance comparison on high dimensional synthetic dataset. $d$ denotes the feature dimension and $H$ is the number of attention heads. }
    \label{fig:hd_pref}
\end{minipage}
\end{figure}

\textbf{Qualitative results and interpretations.} 
Figure~\ref{fig:user_vis} shows the learned user representations given the engagement history. There are three observations. 1) When $H=1$, global query attention fails to capture all the user interests, while the self-attention model is free from the limitation. 2) Viewing from the third row, the self-attention model is more accurate in learning cluster representations than global query models. The latter is systematically biased due to the global query as shown in 
global query models
in Figure~\ref{fig:global_query}.
3) All the models learn super-clusters, depending on the bias in the dataset. For the example shown in Figure~\ref{fig:user_vis}, the two adjacent clusters on the top side of the unit circle are often represented to be a super-cluster.

We also visualize the internal attention scores  and self-attention models (Figure~\ref{fig:self_att_vis}). Some attention heads show highly similar attention patterns because their queries are close to each other, which can be verified from Figure~\ref{fig:global_query}. Figure~\ref{fig:self_att_vis} compares the ground truth attention model with the learned attention. The learned attention shows clear boundaries between clusters in the heatmap. Note that the ground truth ignores the adjacency of clusters but the self-attention model considers the similarity between clusters, so Figure~\ref{fig:self_att_vis}(a) is block diagonal while Figure~\ref{fig:self_att_vis}(b-d) has dark blocks off the diagonal.

\textbf{Quantitative results. }
Previous results show the intuitive comparison between global query models and the self-attention model, and the quantitative results further confirm the consistency of performance gain of self-attention. Experiments are repeated on the dataset for feature dimension $d=16, 32, 64, 128$ and number of attention heads $H=4, 8$. Figure~\ref{fig:hd_pref} shows that the \myname model constantly and significantly outperforms global query models. As illustrated in the 2D dataset, the performance gain benefits from the personalized user representation, rather than matching to the globally popular clusters. Another observation from the result is that for global query models, $H=4$ under-performs $H=8$ models, as the number of attention heads decides the number of global clusters the model can learn; however, for self-attention model, $H=4$ performs even better than $H=8$. The explanation is that the self-attention model does not require a growing number of attention heads with respect to the number of global clusters, and $H=4$ could be already enough for capturing user interest but easier than $H=8$ to train.

\label{ssec:syn_exp}

\subsection{Clustering Options}
\label{ssec:abl_cluster}

\begin{table}[h]
    \centering
    \begin{tabular}{c|c|ccc|ccc}
    \hline \hline 
     \multirow{2}{*}{\makecell{Clustering \\Method}} & 
     \multirow{2}{*}{\makecell{Inference \\ Clusters}} & 
     \multicolumn{3}{c|}{Unweighted \myname} & \multicolumn{3}{c}{ Weighted \myname} \\
     \cline{3-8}
     &  & Amazon & Taobao & MovieLens & Amazon & Taobao & MovieLens \\
     \hline
     None & - & 73.11	& 82.09 & 95.97 & - & - & - \\
     \hline
     \multirow{3}{*}{Ward} & 5 & 71.56	& 80.58 & 95.53 &	79.31 & 8\textbf{6.49} & 94.61 \\
                           & 8 & 71.99 & 80.99 & 95.72 & 80.47 & 87.85 & \textbf{95.25} \\
                           & 10 & 72.16 & 81.20 & \textbf{95.78} &	80.84 & 88.42 & \textbf{95.25} \\
    \hline
    \multirow{3}{*}{K-Means} & 5 & 71.58 &	80.62 &	95.53&	79.26 & 86.18 & \textbf{94.86} \\
                             & 8 & 71.95 & 81.03 & 95.71& 80.66 & 88.02 & 95.17\\
                             & 10 & 72.14 & 81.22 &	95.77&	80.62 & \textbf{88.61} & 95.10\\
    \hline
    \multirow{3}{*}{Spectral} & 5 & \textbf{72.28} & \textbf{80.72} & \textbf{95.54}	& 78.99 & 85.84 & 94.46 \\
                              & 8 & \textbf{72.37} & \textbf{81.08} & \textbf{95.73} & \textbf{80.79} & 87.61 & 94.81 \\
                              & 10 & \textbf{72.64} & \textbf{81.26} & 95.78 & \textbf{81.19} & 88.40  & 95.07\\
    \hline
    \multirow{3}{*}{BIRCH} & 5 & 71.98 & 80.63 & 95.52&	\textbf{79.39} & 86.29 &	94.61\\
                           & 8 & 72.03 &  81.02 & 95.71& 80.65 & \textbf{88.03} & 95.25 \\
                           & 10 & 72.44 & 81.21& 95.78 &80.91 & 88.53 & \textbf{95.25}\\
    \hline
    DBSCAN & - & 71.98 & 80.63 & 95.52 & 70.05 & 75.58 & 89.63 \\
    \hline
    \end{tabular}
    \caption{Comparison of clustering options in AUC (in $10^{-2}$). Note that the number of inference clusters is independent of training, i.e. changing the number of inference clusters does not require the re-training of the model. With the same number of clusters, the best performances are bold.}
    \label{tab:cluster_ablation}
\end{table}

The Ward's algorithm is applied to \myname considering its success in PinnerSage\citep{pinnersage}, it's beneficial to explore the selection of the clustering algorithm and the number of clusters on the collaborative filtering dataset. To illustrate the impact, we evaluate \myname with a wide range of clustering algorithms. 

\textbf{Model Configuration and training}: \myname models are configured with an attention module that takes both positional and temporal encoding. For unweighted \myname, no clustering method is applied to the encoded user engagement history $\{z_*\}$ (computed from Eq.~\ref{eq:compute_z}) in the training stage. For weighted \myname, Ward's algorithm is applied to $\{z_*\}$ and the number of clusters is set to 5. To keep the \myname fully differentiable, the cluster embedding is the encoding of the last item in each cluster, instead of the medoid.  

\textbf{Inference}:
The choice of clustering in the inference phase is independent of its configuration during the training. We explore the inference options on the pre-trained models. Different types of clustering methods are compared: 
\begin{itemize}[noitemsep, topsep=0pt] 
    \item Ward: hierarchical clustering method that minimizes the sum of squared distances within all clusters. 
    \item K-Means: an iterative method also minimizes the sum of in-cluster summed squared distances.
    \item Spectral\citep{spectral_clustering}: performs clustering on the projection of the normalized Laplacian computed from the affinity matrix. 
    \item BIRCH\citep{zhang1996birch}: another hierarchical method that clusters the points by building the Clustering Feature Tree.
    \item DBSCAN\citep{dbscan}: a density-based clustering method that does not require specifying the number of clusters. 
\end{itemize}
The number of clusters is set to 5, 8, and 10 when required. Note that during training, the number of clusters is fixed to 5, however, after training, \myname can produce other numbers of embeddings per user, which gives the system huge flexibility to trade-off between storage/computation cost and recommendation performance. 

\textbf{Result and analysis}: 
There are two observations from Table~\ref{tab:cluster_ablation}. 1) the choice of clustering algorithm has a marginal impact on the performance. While PinnerSage reported that Ward's algorithm outperforms the K-Means, their result does not conflict with our observation here. Recall that for PinnerSage and our experiment on the Pinterest dataset, the clustering method is applied to the exogenous item embeddings, thus the clustering methods can be influenced by the non-flat geometry and outliers. However, with the collaborative filtering dataset, the clustering method is applied to the encodings produced by multi-head self-attention layers which average the embedding of the items and all other items (Eq.~\ref{eq:att}). The encodings after the multi-head self-attention should be smoothly distributed, and as a result, any clustering methods work almost equally well on that. 2) Selecting the number of clusters is a non-trivial trade-off. The motivation to decrease the number of clusters is the storage and computation cost which grow linearly as the number of clusters increases. For unweighted \myname, though the non-clustering (each item is a cluster) settings have the best AUC, decreasing the number of user embedding from 50 (non-clustering) to 10 is still acceptable. For weighted \myname, since it's impossible to learn the clustering weights without applying a clustering method, the trade-off can be more complicated: besides the storage concern, when the number of clusters increases the average information to learn the weights of each cluster decreases, and consequently may hurt the overall performance; on the other hand, 10-cluster settings are better than the 5-cluster settings for all the dataset.

\subsection{Model Latency Comparison}
\label{ssec:latency_cmpr}

\begin{table}[h]
    \centering
    \begin{tabular}{c|c|c|c|c|c|c|c|c}
    \hline\hline
       \makecell{Latency/\\Recall}  &  GRU4Rec & \makecell{BERT4Rec \\ ($L=1$)} & \makecell{BERT4Rec\\ ($L=2$)} & PinText2 & TiSASRec & ComiRec & \makecell{MIP\\(total)} & \makecell{MIP\\(clustering)} \\
       \hline
       Train & \textbf{218.57} & 925.36 &1058.34 & 555.79 & 452.94 & 479.59 & 998.62 & 5.86\\
       (std) & (0.81) & (108.96) & (733.33) & (76.96) & (67.07) & (67.50) & (56.31) & (0.76) \\
       \hline
       Inference & \textbf{1.15} & 38.34 & 57.53 & 14.46 & 14.54 & 14.61 &  40.05 & 5.93 \\
       (std) & (0.20) & (56.60) & (56.23) & (54.99) & (56.34) & (55.70) & (27.08) & (0.72) \\
        \hline
        R@50 & 63.50  & 63.15 & 66.52 & 54.13 & 66.67 & 67.36 & \textbf{78.85} & - \\
        \hline
    \end{tabular}
    \caption{Latency and performance comparison of the models. Training and inference latencies are measured in \emph{ms}, and brackets show the standard deviations.}
    \label{tab:latency_cmpr}
\end{table}

Seeing the performance gain, another prominent question will be what is the time cost of the performance increase. In this section, we profiled the model latency on a desktop computer with a 12-core Intel i7-8700k CPU, and a single Nvidia GeForce RTX 2080 Ti GPU. The neural network training and inference are on the GPU with vanilla PyTorch framework (version 1.12) without any further optimization on the computation. The clustering algorithm in \myname is performed by CPU with Python's scikit-learn package. We set the batch size to 1 and the dataset to Amazon, then measure and summarize the training and inference latency in Table~\ref{tab:latency_cmpr}. 

There are a few observations from Table~\ref{tab:latency_cmpr}. First, compared to the neural network inference latency, the clustering step time cost is trivial. PinText2, ComiRec, and TiSASRec has similar training and inference latency, while the performance of them is worse than \myname. BERT4Rec has similar latency as \myname since our sequential model architecture is similar, while the BERT4Rec has worse performance. GRU4Rec has the least inference time. Notice that the standard deviations of the inference latency of PinText2, TiSASRec, and ComiRec are large. It indicates, though on average the three models are faster in inference, \myname inference latency is less possible to be very large while the other latency might be several times longer than average. 

Conclusively, \myname, as well as other baselines compared, can all satisfy the latency requirement when applying online, even without further optimization on the computation and serving. \myname has a higher time cost compared to some of the baselines, but the performance increase is also appealing.

\end{document}